\documentclass[%
 aip,
 cha,
 amsmath,amssymb,
 reprint,
 numerical,
 twoside
]{revtex4-1}

\usepackage{graphicx}
\usepackage{dcolumn}
\usepackage{bm}
\usepackage[colorlinks=true,linkcolor=blue]{hyperref}%

\usepackage[utf8]{inputenc}
\usepackage{lmodern}
\usepackage{bbold}
\usepackage{amssymb}
\usepackage{color}
\usepackage{amsmath}
\usepackage{bbm}
\usepackage{ifthen}
\usepackage{makecell}
\usepackage[caption=false]{subfig}
\usepackage{multirow}
\usepackage{listings}

\usepackage[ngerman,english]{babel}

\usepackage[absolute]{textpos}
\newcommand{\vect}[1]{
    \ifx#1\eta
        \boldsymbol{#1}
    \else
        \mathbf{#1}
    \fi
}

\newcommand{\mat}[1]{    
    \ifx#1\eta
        \boldsymbol{#1}
    \else
        \mathbf{#1}
    \fi
}

\newcommand{\D}[0]{
    \mathrm{D}
}

\newcommand{\idx}[1]{
    \mathrm{#1}
}

\newcommand{\diff}[1]{
     \mathrm{d} #1 
}

\newcommand{\var}{
     \operatorname{Var}
}

\newcommand{\std}{
     \operatorname{Std}
}

\newcommand{\hateq}{
     \hat{=} 
}

\newcommand{\T}{
    ^\mathrm{tr}
}

\newcommand{\NP}{
    {N_P}
}

\begin{document}

\title{Basin structure of optimization based state and parameter estimation }

\author{Jan Schumann-Bischoff}
\affiliation{ Biomedical Physics Group, 
        Max Planck Institute for Dynamics and Self-Organization, 
        Am Faßberg 17, 37077 Göttingen, Germany}
\affiliation{Institute for Nonlinear Dynamics, Georg-August-Universität 
        Göttingen, Am Faßberg 17, 37077 Göttingen, Germany}

\author{Ulrich Parlitz}
\email{ulrich.parlitz@ds.mpg.de}
\affiliation{ Biomedical Physics Group, 
        Max Planck Institute for Dynamics and Self-Organization, 
        Am Faßberg 17, 37077 Göttingen, Germany}
\affiliation{Institute for Nonlinear Dynamics, Georg-August-Universität 
        Göttingen, Am Faßberg 17, 37077 Göttingen, Germany}
  
\author{Henry D. I. Abarbanel}
\email{habarbanel@ucsd.edu}
\affiliation{Department of Physics, University of California, San Diego,
        9500 Gilman Drive, La Jolla, CA 92093-0374}
\affiliation{Marine Physical Laboratory (Scripps Institution of Oceanography)}

\author{Mark Kostuk}
\author{Daniel Rey}
\author{Michael Eldridge}
\affiliation{Department of Physics, University of California, San Diego,
        9500 Gilman Drive, La Jolla, CA 92093-0374}

\author{Stefan Luther}
\email{stefan.luther@ds.mpg.de}
\affiliation{ Biomedical Physics Group, 
        Max Planck Institute for Dynamics and Self-Organization, 
        Am Faßberg 17, 37077 G\"ottingen, Germany}
\affiliation{Institute for Nonlinear Dynamics, Georg-August-Universität 
        Göttingen, Am Faßberg 17, 37077 G\"ottingen, Germany}
\date{\today}

\begin{abstract}
Most data based state and parameter estimation methods require suitable 
initial values or guesses to achieve convergence to the desired solution, which 
typically is a global minimum of some cost function. Unfortunately, however, 
other stable solutions (e.g., local minima) may exist and provide suboptimal or 
even wrong estimates. Here we demonstrate for a 9-dimensional 
Lorenz-96 model how to characterize the basin size of the global minimum 
when applying some particular optimization based estimation algorithm. We 
compare three different strategies for generating suitable initial guesses and we
investigate the dependence of the solution on the given trajectory segment 
(underlying the measured time series). To address the question of how many state 
variables have to be measured for optimal performance, different types of 
multivariate time series are considered consisting of 1, 2, or 3 variables.
Based on these time series the local observability of state 
variables and parameters of the Lorenz-96 model is investigated 
and confirmed using delay coordinates.  
This result is in good agreement with the observation  that correct state and 
parameter estimation results are obtained if the  
optimization algorithm is initialized with initial guesses close to the true solution.  
In contrast, initialization with other exact solutions of the model equations 
(different from the true solution used to generate the time series) typically fails, 
i.e. the optimization  procedure ends up in local minima different from the true solution.
Initialization using random values in a box around the attractor exhibits success rates 
depending on the number of observables and the available time series (trajectory 
segment).
\end{abstract}

\maketitle

\begin{quotation}
For many physical processes dynamical models are available but often not all 
their state variables and (fixed) parameters are known or easily accessible. In meteorology, 
for example, sophisticated large scale models exist, which have to be 
continuously adapted to the true temporal changes of temperatures, wind speed, 
humidity, and other relevant physical quantities. In quantitative biology
mathematical models of single neural or cardiac cells or networks may contain 
many state variables and parameters whose values are not easy to measure (without 
destroying the system). In such cases,  data based estimation 
methods can be used  to determine these unknown states and a parameters by 
adapting a suitable model to reproduce and predict the measured time series. 
This approach can be successful only if two conditions are fulfilled:
(i) the available data have to provide sufficient information, i.e. the unknown state variables 
and parameters have to be \textit{observable}
and (ii) the estimation algorithm has to be properly initialized with initial guesses 
sufficiently close to the true solution. Here, we consider both problems for the 
Lorenz-96 model and compare different initialization methods in terms of their effective basin sizes. 
\end{quotation}

\section{Introduction}       \label{sec:intro}
%
\begin{textblock}{180}(18,262)
\noindent Copyright 2015 American Institute of Physics. This article may be 
downloaded for personal use only. Any other use requires prior permission of the 
author and the American Institute of Physics. The following article appeared in 
\newline J. Schumann-Bischoff \textit{et al.}, Chaos \textbf{25}, 053108 (2015) 
and may be found at \url{http://dx.doi.org/10.1063/1.4920942}.
\end{textblock}
Estimation methods for state variables or (fixed) parameters can be implemented 
employing 
synchronization \cite{PJK96,PKK98,JP00,SRL09,REKASBP14} or 
optimization methods \cite{CGA08,B10,SBP11,SBLP13}, for example. 
In the literature one can find many examples with successful applications of 
state and parameter estimation methods even for 
chaotic systems \cite{VTK04,ACFK09,QA10,A13,YKRAQ15}. 
In practice, however, attempts to fit a model 
(for example, a set of nonlinear ordinary differential equations (ODEs)) to 
given data may fail. There are many possible 
reasons for such a failure, including inappropriate models, poor quality of the measured 
time series (too noisy, too short), or external perturbations not covered by the 
model. But even with relatively clean data and the right model architecture, 
estimation may turn out to be difficult, because the available data do not 
contain sufficient information about the underlying process. 
Therefore, in this article we address how the success of a given estimation 
algorithm for a given model depends of the following aspects:
\begin{itemize}
    \item[(a)]  the number of available observables (in a multivariate time  series)
    \item[(b)]  the available time series (corresponding to some particular 
                trajectory segment)
    \item[(c)]  and the way the estimation algorithm is initialized (using guesses for the unknown quantities).
\end{itemize}
The focus in the presented analysis is on models given by ODEs,
\begin{align}
    \frac{\diff{\vect{x}(t)}}{\diff{t}} &=
    \vect{F}(\vect{x}(t),\vect{p}, t) \, ,
    \label{eq:ode_model}
\end{align} 
and a measurement function,
\begin{align}
    \vect{y}(t)     &= \vect{h}(\vect{x}(t)) \in \mathbb{R}^L\, ,
    \label{eq:measurementfct}
\end{align}
representing the model output
with a state vector $\vect{x}(t) = (x_1(t),\dots , x_D(t))\T \in 
\mathbb{R}^D$, and model parameters $\vect{p} = (p_1,\dots ,p_\NP)\T \in 
\mathbb{R}^\NP$. Here and in the following the superscript ``$\mathrm{tr}$'' 
denotes the transpose.  We assume that a multivariate $L$-dimensional 
(experimental) time series $\{\vect{\eta}(n)\}$ is given consisting of $N+1$ 
samples $\vect{\eta}(n) \hateq \vect{\eta}(t_n) \in \mathbb{R}^L$, analogous to 
the model output, and measured at times ${\cal{T}} = \{ t_n = n\cdot \Delta t 
\mid n = 0,1, \dots, N\}$.
The observation times $t_n$ are equally spaced (with a fixed time
step $\Delta t$) and start at $t_0 = 0$.
Any solution of Eq.~\eqref{eq:ode_model} at discrete times $t_n = n \cdot 
\Delta t$ with a fixed 
time step $\Delta t$ is denoted by $\{\vect{x}(n)\}$ and consists of $N+1$ 
samples $\vect{x}(n) \hateq \vect{x}(t_n)$ at
times $t_n \in {\cal{T}}$. If, from the context, $\Delta t$ and the range of $n$ 
are clear, this information will be dropped in the following. The same 
convention holds if a solution is denoted by another symbol, for example  
$\vect{z}$ instead of $\vect{x}$: the solution $\{\vect{z}(n)\}$ consists of 
$N+1$ samples $\vect{z}(n) \hateq \vect{z}(t_n)$ measured at
times $t_n \in {\cal{T}}$.

As an example we use synthetic data from a 9-dimensional Lorenz-96 
system \cite{Lorenz96} (Sec.~\ref{sec:l96_example}) and an optimization based 
estimation 
algorithm \cite{SBP11} (Sec.~\ref{sec:estimalg}). We check the observability  of 
the state variables of the Lorenz-96 model which are not ``measured'' (i.e., not 
contained in the multivariate time series) using a local analysis 
(Sec.~\ref{sec:local_observability}) employing the Jacobian matrix of the delay coordinates 
map \cite{PSBL14_PRE,PSBL14_Chaos}. This analysis  indicates that even with a 
single  observable (scalar time series) all state variables and the parameter of the 
Lorenz-96 system are in principle (locally) observable. 

To investigate \textit{global convergence}  features 
(using initial guesses that are not close to the true solution)
we probe the basin structure of the observability problem 
by considering 18 different trajectories of the 
Lorenz-96 system (on the same chaotic attractor, but generated with different 
initial conditions). From each trajectory 15 different (multi-variate) time 
series are derived consisting of one, two, or three observables. Then a
particular method for generating initial guesses to initialize the 
optimization algorithm is chosen, and the 
estimation algorithm is applied to each of these 15 time series 500 times (with 
different random initial guesses) to obtain statistics of how often the 
estimation problem is solved successfully. 
In other words, we compute the probability that a generated initial guess is 
located in the basin of the true solution of the given optimization algorithm.
This method of estimating the ``basin size'' was adopted from Menck et al. \cite{MHMK13}.

In Sec.~\ref{sec:l96_example} we introduce the Lorenz-96 model
which will serve as an example for the following studies. First, local observability of the state 
variables and the parameter of the Lorenz-96 model  is investigated and confirmed in 
Sec.~\ref{sec:local_observability}.  Then, in Sec.~\ref{sec:estimalg} we present 
the estimation algorithm used and in Sec.~\ref{sec:determining_basin_size} our approach for 
characterizing the size of the basin of the true solution is introduced. The 
true solution is  a stable fixed point of the optimization algorithm with a basin of attraction 
and the desired estimation of the true solution is only possible if the optimization algorithm 
is initialized  with guesses from this basin. 
To check the stability  of this fixed point the  optimization procedure was initialized by 
initial guesses consisting of randomly perturbed  true values. For all these initial guesses 
the optimization results converged to the true solution.
However, since in general the location of the true solution is not known the 
size and the structure of its basin are most important for any initialization  strategy. 
Three possible initialization methods 
(Sec.~\ref{sec:determining_basin_size}B) are 
invested in detail and compared in terms of their efficacy for finding the true solution. 
All results are summarized in the conclusion drawn in 
Sec.~\ref{sec:discussion_and_conclusion}.

\section{Example: The Lorenz-96 model}    \label{sec:l96_example}
%
As an example for demonstrating the proposed analysis 
we consider in the following a $D=9$ dimensional Lorenz-96 model
\begin{align}
    \frac{\diff{x_i(t)}}{\diff{t}}  &= x_{i-1}(t) \cdot 
        (x_{i+1}(t)-x_{i-2}(t)) - x_i + p 
    \label{eq:lorenz96_model}
\end{align}
with $p=8.17$ and a cyclic index $i$  ($x_{D+1}(t) = x_1(t)$, 
$x_{0}(t) = x_D(t)$, and $x_{-1}(t) = x_{D-1}(t)$). 
For the parameter value $p=8.17$ the model generates a chaotic attractor.

The Lorenz-96 model is chosen here as an example because previous 
investigations showed that it is very difficult to estimate its state variables 
and the parameter $p$ using only a few observables\cite{PhD_Kostuk2012}.
Recently, however, Rey et. al. \cite{REKASBP14} demonstrated  successful
state and parameter estimation based on univariate time series consisting of a 
single Lorenz-96 state variable and a  synchronization scheme employing delay coordinates. 
Law et al. \cite{LSSS14}  applied the extended Kalman filter 
and the 3D-VAR data assimilation technique to the chaotic Lorenz-96 model and 
also encountered difficulties in the estimation of model state variables if only 
few model state variables are observed. 

Technically, the Lorenz-96 model \eqref{eq:lorenz96_model} is used here in
a twin experiment for both, (i) generating the ``measured'' time series and (ii) as a model to be 
adapted to a (multivariate) time series using the optimization based estimation method
described in Sec. \ref{sec:estimalg}. 

To address the question how many observables have to be known for successful 
state and parameter estimation we consider multivariate time series $\{ \vect{\eta}(n) \}$ with one,
two, or three  state variables.
More precisely, for the 9 dimensional Lorenz-96 model we consider \textit{all} possible 
combinations of one to three state variables as being ``measured''. For example, let 
us assume we can measure the state variables $(x_1, x_2, x_5)$. Due to the  symmetry
in Eq.~\eqref{eq:lorenz96_model}, sampling $(x_1,x_2,x_5)$ is equal
to measuring $(x_3,x_4,x_7)$ or $(x_7,x_8,x_2)$. Hence checking the
observability of all state variables and the parameter $p$ with the given 
multivariate time series $(x_1,x_2,x_5)$ is equivalent to
checking the observability with the time series $(x_7,x_8,x_2)$. 
Removing all mathematically equivalent combinations results in the 
following 15 distinct combinations of state variables: $x_1$, $(x_1,x_2)$, 
$(x_1,x_3)$,
$(x_1,x_4)$, $(x_1,x_5)$,
$(x_1,x_2,x_3)$, $(x_1,x_2,x_4)$, 
$(x_1,x_2,x_5)$, $(x_1,x_2,x_6)$, $(x_1,x_2,x_7)$, $(x_1,x_2,x_8)$,
$(x_1,x_3,x_5)$, $(x_1,x_3,x_6)$, $(x_1,x_3,x_7)$, and  
$(x_1,x_4,x_7)$.

\section{Local observability of model state variables and fixed parameters}     \label{sec:local_observability} 
%
We consider models given by a set of $D$ coupled ODEs, 
Eq.~\eqref{eq:ode_model}, with a measurement function, 
Eq.~\eqref{eq:measurementfct}, representing the relation between model states 
$\vect{x}(n)$  and the model output $\vect{y}(n)$ corresponding to the 
observations $\{\vect{\eta}(n)\}$.
The state vector(s) $\vect{x}(t)$ and the model parameters $\vect{p}$
are unknown and have to be estimated from a (multivariate) time series.
The technique used in this article to adapt a model given by ODEs 
\eqref{eq:ode_model} to a (multivariate) time 
series given by $\{\vect{\eta}(n)\}$ with a measurement function 
\eqref{eq:measurementfct} will be described in Sec.~\ref{sec:estimalg}. 
Similar to other methods for state and parameter estimation, 
this algorithm will provide estimates for the model state variables and the (fixed) model
parameters (except if, for example, numerical problems arise).
The fact, however, that an algorithm produces some output does not mean 
that this output is correct or useful.
Therefore, a method is needed which indicates whether it is (in principle) possible to 
estimate $\vect{p}$ and $\vect{x}(t)$ correctly from $\{\vect{\eta}(n)\}$.
This question addresses the general problem of observability, 
which is well known  from control  theory \cite{LM67,HK77,NS90,Sontag,LAM05}. 
In the following section we shall employ \textit{the time delay coordinates map} of the 
observed time series to investigate local observability following an approach presented 
in Refs. \cite{PSBL14_PRE,PSBL14_Chaos}.

\subsection{The delay coordinates map}
%
Let the dynamical system \eqref{eq:ode_model} generate a flow 
\begin{align} 
  \phi^\tau : \mathbb{R}^D \otimes \mathbb{R}^\NP   &  \rightarrow  
                \mathbb{R}^D \\ \nonumber
              (\vect{x}(t), \vect{p}) & \mapsto   \vect{x}(t+\tau)
\end{align}
mapping a state $\vect{x}(t)$ at time $t \in \mathbb{R} $ to a (future) state $\vect{x}(t+\tau)$.
Furthermore, delay coordinates are given via the $L$ dimensional 
measurement function Eq.~\eqref{eq:measurementfct}
\begin{equation}
      \vect{y}(t+\tau) = \vect{h}( \vect{x}(t+\tau) )  = \vect{h}( 
                            \phi^{\tau}(\vect{x}(t), \vect{p})) 
\end{equation}
from a trajectory starting at $\vect{x}(t)$ with delay time 
$\tau$. If the delay time is $\tau = 0$, then we obtain
$\phi^{0}(\vect{x}(t), \vect{p}) = \vect{x}(t)$ and recover  $\vect{y}(t) = 
\vect{h}(\vect{x}(t))$.
Taking into account $K$ time steps we can define a $D_M = K \cdot L$ 
-dimensional \textit{delay coordinates map}
\begin{align}
    \vect{s} &= \vect{S}( \vect{x}(t), \vect{p} )  \notag \\
             &= \left(\vect{y}\T(t),  \vect{y}\T(t+\tau), \dots ,  
                \vect{y}\T(t+(K-1)\tau) \right) .
    \label{eq:G}
\end{align}
Here the delay coordinates map $\vect{S}$ is considered as  a function of: 
(i) the state $\vect{x}(t)$ and the parameters $\vect{p}$ of the underlying system, and 
(ii) of the delay time $\tau$ (not listed as an argument of $\vect{S}$ here, 
because $\tau$ is fixed and not part of the estimation problem). 
All $\vect{y}\T (t+i\cdot \tau)$, $i=0, \dots, K-1$ are 
row vectors. Hence, the right hand side of Eq.~\eqref{eq:G} is a (row) vector 
containing $K \cdot L$ elements.

If the delay coordinates map Eq.~\eqref{eq:G}, 
$\vect{S}: \mathbb{R}^D \otimes \mathbb{R}^\NP \rightarrow \mathbb{R}^{D_M}$, 
is locally  
invertible, then the full state $\vect{x}(t)$ and the parameter vector $\vect{p}$ 
can be uniquely determined from the signal $\vect{h}(\vect{x}(t))$, which, in 
a real world experiment, corresponds to the measured time series 
$\{ \vect{\eta}(t_n) \}$ Eq.~\eqref{eq:measurementfct}. 
Mathematically, the delay coordinates map Eq.~\eqref{eq:G} has to be an immersion \cite{Sauer},
i.e. the Jacobian matrix $\D_{\vect{x},\vect{p}} \vect{S} = 
\D_{\vect{x},\vect{p}} 
\vect{S}(\vect{x}(t),\vect{p})$  of the delay coordinates map $\vect{S}$ has to 
have maximal (full) rank.

The accuracy and robustness of estimated state variables or fixed parameters can be quantified by 
the \textit{uncertainty} 
\begin{equation} 
      \nu_j = \sqrt{ \left[ \D_{\vect{x},\vect{p}}\vect{S}\T \cdot 
\D_{\vect{x},\vect{p}}\vect{S}  \right]^{-1}_{jj} } 
      \label{eq:nu}
\end{equation}
of state variables ($j=1,\ldots, D$) and parameters ($j=D+1, \ldots, D+\NP$) which 
was introduced in Ref.\cite{PSBL14_PRE,PSBL14_Chaos}. Perturbations of the 
measured time series are amplified by $\nu_j$, i.e. the larger $\nu_j$ the less 
precise is the estimation of the corresponding state variable or fixed parameter. Note that 
the uncertainty $\nu_j$ depends (via $\D_{\vect{x},\vect{p}}  
\vect{S}(\vect{x}(t),\vect{p})$) on the location in state and parameter space.

\subsection{Local observability of the Lorenz-96 model}     \label{sec:l96_local_observability}
%
To assess the local observability of the states $\vect{x}(t)$ and the (single) model 
parameter $p$ of the $D=9$ dimensional Lorenz-96 model \eqref{eq:lorenz96_model}
their uncertainty is checked at $10^4$ arbitrary reference points on the 
attractor. To obtain  the reference points the Lorenz-96 model was integrated 
$10^7$ steps with a  step size of 0.01 using a Runge-Kutta-45 integration 
scheme. Then every 1000th point was picked as a reference point $\mathbf{x}(t)$ 
for the observability analysis described in the following.

As mentioned in Sec. \ref{sec:l96_example}  for the 9-dimensional Lorenz-96 model 
there are 15 different combinations of one to three state variables constituting a multivariate time series. 
We select the following two cases as representative examples:
\begin{itemize}

\item[(a)]  measurement function $h(\vect{x}(t))=x_1(t)$  (i.e., $L=1$) with 
            $K=12$ and a resulting delay reconstruction dimension of $D_M=12$ 
            and 

\item[(b)]  measurement function $\vect{h}(\vect{x}(t))=(x_1(t),x_3(t),x_6(t))$ 
            (i.e., $L=3$) with $K=4$ and hence a delay reconstruction 
            dimension of $D_M=12$  (see Eq.~\eqref{eq:G}).
\end{itemize}

\begin{figure*} 
    \centering
    \includegraphics[width=1\textwidth]{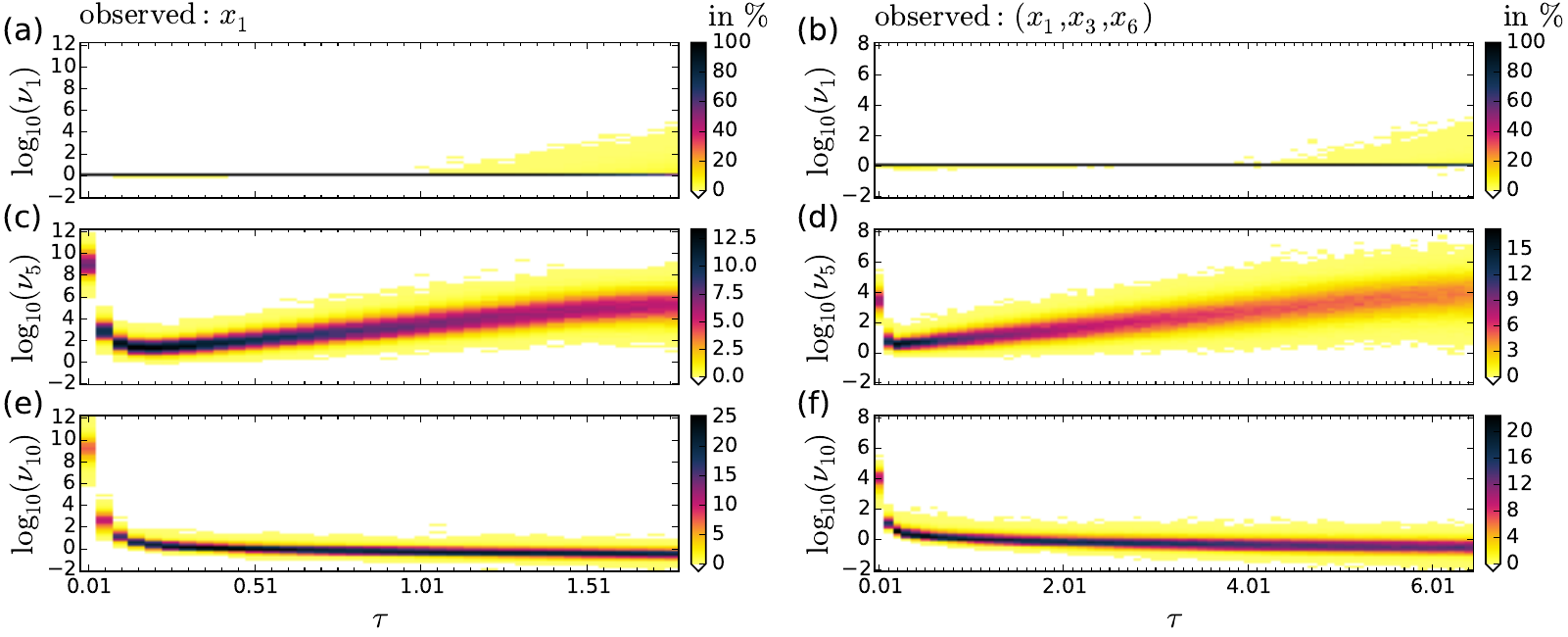}
    \caption{(Color online) Probability distributions (color coded frequency in 
            \%, vertically plotted) of 
            uncertainties $\nu_1$, $\nu_5$ and $\nu_{10}$ vs. delay time $\tau$ 
            for the Lorenz-96 model. The uncertainties $\nu_1$ and $\nu_5$ 
            correspond to state variables $x_1$ and $x_5$, respectively,
            and $\nu_{10}$ corresponds to the parameter $p$. In the left column  
            ((a),(c),(e)) a scalar ($L=1$ dimensional) measurement function  
            $h(\vect{x}(t))=x_1(t) $ is used for generating delay coordinates with 
            $K=12$. The right column ((b),(d),(f)) shows results for 
            the $L=3$ dimensional measurement function 
            $\vect{h}(\vect{x}(t))=(x_1(t),x_3(t),x_6(t))$ and a delay 
            reconstruction (Eq.~\eqref{eq:G}) with $K=4$.
            Hence $D_M=12$ dimensional delay coordinates 
            (Eq.~\eqref{eq:G}) are constructed in both cases. 
            The histograms for $\nu_1$ are very similar to histograms obtained 
            for other measured state  variables ($x_3$, $x_6$ in the right column), not 
            shown here. Similarly, the histograms of $\nu_5$ are representative 
            for histograms of the remaining unmeasured state variables (not shown 
            here).}
    \label{fig:fig1_lorenz96_nu_vs_tau}
\end{figure*}

The reconstruction dimension $D_M = L \cdot K$  is the same in both cases.
Histograms of the uncertainties $\nu_j$, for $j=1,\dots, 10$, Eq.~\eqref{eq:nu}, are 
computed for the $10^4$ reference points on the attractor. 
Figure \ref{fig:fig1_lorenz96_nu_vs_tau} shows the 
histograms for $\nu_1$, $\nu_5$ and $\nu_{10}$
which are plotted vertically using color coding (relative frequencies 
of the corresponding uncertainties are given in percent, see color bar).
All distributions shown here are unimodal.
%
The left column shows the results for $h(\vect{x}(t))=x_1(t)$ and the right 
column for $\vect{h}(\vect{x}(t))=(x_1(t),x_3(t),x_6(t))$. In both cases the uncertainty
$\nu_1$ corresponds to the ``measured'' state variable $x_1$, $\nu_5$ corresponds to 
the ``hidden'' state variable $x_5$ (not measured directly), and $\nu_{10}$ corresponds to the model 
parameter $p$. Histograms for the other $\nu_i$ are not shown, because they 
look very similar to the histograms for $\nu_1$ (if the corresponding 
state variable is measured) and $\nu_5$ (if the corresponding model state variable is 
unmeasured). The reason for these similarities is the symmetry in the Lorenz-96 
model equations 
\eqref{eq:lorenz96_model}. 

For both measurement functions one can see that the uncertainty values
of the maxima of the histograms exhibit a U-shaped dependence on the delay time $\tau$.
The smallest uncertainties occur between $\tau=0.11$ and $\tau=0.21$ for the 
\textit{unmeasured} state variables 
(Figs.~\ref{fig:fig1_lorenz96_nu_vs_tau}c,d). In both 
cases the distribution of the uncertainty $\nu_1$  of  the \textit{measured}
state variable  (and similar for $\nu_3$ and $\nu_6$ if $R=3$, not shown)  
possess a sharp peak around $\log_{10} \nu_i = 0$.
In this case the state variable $x_1$ is an observed quantity, and therefore, 
the delay reconstruction does not provide much further information about its values. 
For relatively large delay times ( $\tau >1$ in (a) and  $\tau > 4$ in (b)) the delay 
reconstruction becomes rather  ``poor''. The measurement noise is amplified resulting in 
a tail of the histogram with uncertainty values larger than one (the yellow areas above
the horizontal line at $ \log(\nu_1)=0$). This result is not surprising, because in nonlinear 
time series analysis it is well known that the delay time has to be chosen carefully and 
must not be too large for chaotic attractors,
since otherwise  the reconstructed attractor will be heavily distorted.
In contrast, the $\nu$-values of the centers of the distributions of the 
uncertainty  $\nu_{10}$ of the parameter $p$ 
decrease with increasing $\tau$ until the maximum of the distribution is below one
(see Figs.~\ref{fig:fig1_lorenz96_nu_vs_tau}e,f)  (without exhibiting a  clear 
minimum).

As mentioned before, these two examples are representative for all 
multivariate time series from the Lorenz-96 model consisting of combinations 
of one to three  state variables. The other 13 combinations show 
similar histograms. If a certain state variable in one of the other combinations is 
measured, then the corresponding histogram of the corresponding uncertainty
looks similar to the histograms 
of $\nu_1$ in Fig.~\ref{fig:fig1_lorenz96_nu_vs_tau}a. The same holds for 
state variables that are not contained in the multivariate time series (and where 
the corresponding $\nu$ histograms look similar to the 
histograms of $\nu_5$ in Fig. \ref{fig:fig1_lorenz96_nu_vs_tau}) and the model 
parameter $p$ (the corresponding $\nu_{10}$ histograms look similar to the 
histograms of $\nu_{10}$ in Fig. \ref{fig:fig1_lorenz96_nu_vs_tau}). With an
increasing number of measured  state variables (from one to three) the maxima of 
the distributions at the minimum of the histograms of $\nu$ for unmeasured 
state variables move only slightly in the direction of smaller $\nu$ (see, for example,
Figs.~\ref{fig:fig1_lorenz96_nu_vs_tau}c,d).   This trend 
also holds for all combinations of four measured state variables (not shown here). 
The fact that the histograms for one to three observed state variables look almost 
the same means that the \textit{local} observability of the model 
parameter and the unmeasured state variables is maintained, even if only a 
\textit{single} model state variable is measured instead of three (or more).
In the following we shall investigate whether this local result holds 
for states that are not close to the true solution. Specifically, we want to know if 
we can uniquely recover the true solution using a univariate time series 
(from a single observable), even if the initial guesses are far from the true solution. 
Since the true solution is a stable fixed point of any optimization based estimation algorithm, 
we are thus interested in the 'size'  and the structure of the basin of attraction of 
this fixed point.   To address this question we shall introduce
in the next section a particular estimation algorithm which is used herev as 
a prototypical example for investigating the solution basin.

\section{State and Parameter estimation algorithm}     \label{sec:estimalg}
%
The method used in this study to adapt a model to a time series is based on
minimizing a cost function and in the context of this paper it represents only 
one out of many different state and parameter estimation methods 
\cite{CGA08,B10} where the same type of basin size analysis could be applied.
The method  was introduced in Ref. \cite{SBP11} and will be summarized in the 
following.

The goal of the estimation process is to find a set of values for the model state
variables $\vect{x}(t)$ at each time step of its discretization and the model 
parameters $\vect{p}$ such that the model equations, given by a set of ODEs 
(see Eq.~\eqref{eq:ode_model}), provide via the measurement function 
\eqref{eq:measurementfct} a model times series $\{ \vect{y}(n) \}$ consisting
of $N+1$ samples $\vect{y}(n) \hateq \vect{y}(t_n) \in \mathbb{R}^L$ with $t_n 
\in {\cal{T}} \, \forall \, n$
that matches the experimental time series $\{\vect{\eta}(n) \}$. In other 
words, the average difference between $\vect{\eta}(n)$ and $\vect{y}(n)$ 
should be small. 

Furthermore, the model equations should be fulfilled as well as possible. This 
means that modeling errors $\vect{u}(t)$ are allowed, but 
should be small. Therefore, model \eqref{eq:ode_model} is extended to include 
modeling errors $\vect{u}(t)$,
\begin{align}
    \frac{\diff{\vect{x}(t)}}{\diff{t}} &=
\vect{F}(\vect{x}(t),\vect{p}, t) + \vect{u}(t) \, ,
    \label{eq:modelu}
\end{align} 
so that when $\vect{u}(t)$ is small the model trajectory $\vect{x}(t)$ closely 
matches the model equations. To incorporate model error into the optimization, 
we discretize $\vect{u}(t)$ and $\vect{x}(t)$ at times $t_n \in {\cal{T}}$ so 
that the state variables $\{ \vect{x}(n) \}$, $\vect{x}(n)  \hateq 
\vect{x}(t_n)$, at each time $n$ must be estimated in addition to $\vect{p}$. 
For simplicity, we choose $\vect{x}(t)$ and $\vect{y}(t)$ to be sampled at the 
same times that the data are observed.
Similarly, $\vect{u}(t)$ is discretized to $\{\vect{u}(n)\}$ with 
 $\vect{u}(n) \hateq \vect{u}(t_n)$ and $t_n \in {\cal{T}}$.
The discretization of \eqref{eq:modelu} is given by
\begin{align}
    \vect{u}(n) &\approx \left. \frac{\Delta \vect{x}}{\Delta t} \right|_{t_n}
                            - \vect{F}(\vect{x}(n), \vect{p},t_n)  \, ,
    \label{eq:findiffmodel}
\end{align}
where the symbol $\left. \frac{\Delta \vect{x}}{\Delta t} \right|_{t_n}$
stands for the finite difference approximation of 
$\frac{\diff{\vect{x}(t)}}{\diff{t}}$ at time $t_n$. 

The goal of the adaption
process is to minimize (on average) the norm of $\vect{u}(n)$ \textit{and} the
norm of the difference $\vect{\eta}(n) - \vect{y}(n)$ for all $n \in {\cal{T}}$.
Technically, this optimization problem can be implemented in different ways \cite{ACFK09}
and in the following we use unconstrained optimization \cite{SBP11} employing 
automatic differentiation  \cite{SBLP13}. The cost function used in this study 
can be derived from a general probabilistic description of the estimation 
problem assuming Gaussian distributions (also called weakly constrained 4D-VAR 
in geosciences)  \cite{E97,LE96,QA10,Evensen09} and consists of four terms
\begin{align}
    C(\{ \vect{x}(n) \},\vect{p}) &= C_1 + C_2 + C_3 + C_4
    \label{eq:cost}
\end{align}
with
\begin{align}
    C_1 &=  \frac{\alpha}{N+1}\cdot\sum_{n=0}^N\left(\vect{\eta}(n)
-            \vect{y}(n)\right)\T \vect{A} \left(\vect{\eta}(n) -
            \vect{y}(n)\right) 
            \label{eq:ts_model_diff} \\
    C_2 &=  \frac{1-\alpha}{N+1}\cdot\sum_{n=0}^N \vect{u}(n)\T \vect{B}
            \vect{u}(n) 
            \label{eq:modellingerr} \\
    C_3 &=  \frac{1-\alpha}{N+1}\cdot\sum_{n = 3}^{N-2}
            \left(\vect{x}_\idx{apr}(n)-\vect{x}(n)\right)\T \vect{E}
            \left(\vect{x}_\idx{apr}(n)-\vect{x}(n)\right) 
            \label{eq:hermite4cost} \\
    C_4 &=  \beta\cdot\vect{q}
            (\vect{w},\vect{w}_\idx{l},\vect{w}_\idx{u})\T \cdot     
            \vect{q}(\vect{w},\vect{w}_\idx{l},\vect{w}_\idx{u}) \, .
            \label{eq:costbounds}
\end{align}
The term $C_1$ penalizes the difference between $\vect{\eta}(n)$ and
$\vect{y}(n)$ whereas $C_2$ penalizes large magnitudes of $\vect{u}(n)$. 
In the term $C_3$ a
Hermite interpolation is performed to determine $\vect{x}_\mathrm{apr}(n)$
from neighboring points and the time derivatives which are,
according to \eqref{eq:modelu}, given by  $ \vect{F}(\vect{x}(t), \vect{p}, t) + \vect{u}(t) $ 
and provide the approximate solutions
 \begin{align}
    \vect{x}_\idx{apr}(n)  
    = & \frac{11}{54} \left[ \vect{x}(n-2)+ \vect{x}(n+2) \right] +   
        \frac{8}{27} \left[ \vect{x}(n-1) \right. \notag \\
     &  \left. + \vect{x}(n+1)\right] + \frac{\Delta t}{18}   
        \left[\vect{F}(\vect{x}(n-2),\vect{p},t_{n-2}) \right. \notag \\
     & \left. + \vect{u}(t_{n-2}) - \vect{F}(\vect{x}(n+2), \vect{p},t_{n+2})   
              - \vect{u}(t_{n+2}) \right] \notag \\
    &   + \frac{4\Delta t}{9} \left[
        \vect{F}(\vect{x}(n-1), \vect{p},t_{n-1}) + \vect{u}(t_{n-1}) 
        \right. \notag \\
    & \left. - \vect{F}(\vect{x}(n+1), \vect{p},t_{n+1})   - \vect{u}(t_{n+1})
                       \right] \, .
        \label{eq:hermite4}
\end{align} 
Smoothness of $\{ \vect{x}(n) \}$  is enforced by
small differences $\vect{x}_\idx{apr}(n)-\vect{x}(n)$.
The term $C_3$ suppresses non-smooth (oscillating) solutions which
may occur without this term in the cost function. 
In this paper the weight matrices $\mat{A}$, $\mat{B}$ and $\mat{E}$ are 
diagonal matrices. The diagonal elements can be used for an individual 
weighting.

The solution $(\{ \hat{\vect{x}}(n) \},\hat{\vect{p}})$ obtained through the 
optimization of the cost function \eqref{eq:cost} is taken to be the maximum 
likelihood estimate.

Let
\begin{align}
    \vect{w} &= (\{ \vect{x}(n) \}, \vect{p})
    \label{eq:w}
\end{align}
be a vector containing all quantities to be estimated. To force
$\vect{w}$ to stay between the lower and upper bounds $\vect{w}_\idx{l}$ and
$\vect{w}_\idx{u}$, respectively, the vector valued function
$\vect{q}(\vect{w},\vect{w}_\idx{l},\vect{w}_\idx{u}) = (q_1,\dots ,q_L)\T$
is defined as
\begin{align}
    q_i(w_i,w_{\idx{l},i},w_{\idx{u},i}) &= \begin{cases}
        w_{\idx{u},i}-w_\idx{i}     & \mathrm{for} \enskip w_i \geq 
w_{\idx{u},i}
\\
    0                       & \mathrm{for} \enskip w_{\idx{l},i} < w_\idx{i} <
w_{\idx{u},i} \\
      w_{\idx{l},i}-w_\idx{i}   & \mathrm{for} \enskip w_i \leq w_{\idx{l},i}  
.\\
                                \end{cases}     
\end{align}
$q_i$ is zero if the value of $w_i$ lies within its bounds. To enforce this,
the positive parameter $\beta$
is set to a large number, e.g. $10^5$.  

The homotopy parameter $\alpha$ can be
used to control whether the solution should be close to data ($\alpha \approx 1$)
or has a smaller error in fulfilling the model equations.
In Ref. \cite{BS11} a technique is described to find an optimal $\alpha$.
Furthermore, one might use continuation (see Ref. \cite{B10}) where $\alpha$ is stepwise
decreased. Starting with $\alpha \approx 1$ results in a solution close to the
data. Then, $\alpha$ is slightly decreased and the previously obtained
solution is used as an initial guess to optimize the cost function
again. This procedure is repeated until the value $\alpha=0.5$ is reached.

Note that the cost function can be written in the form
\begin{align}
    C(\vect{w}) = \sum_{j=1}^J H_j(\vect{w})^2 
                = \left\Vert \vect{H}(\vect{w}) \right\Vert_2^2 
    \label{eq:costopt}
\end{align}
where $\vect{H}(\vect{w})$ is a high dimensional vector valued function of
the high dimensional vector $\vect{w}$. To optimize \eqref{eq:costopt} we use an
implementation of the Levenberg-Marquardt algorithm \cite{L44,M63}
called \texttt{sparseLM} \cite{sparseLM}. Although $C(\vect{w})$ will be
optimized, \texttt{sparseLM} requires $\vect{H}(\vect{w})$ and the sparse
Jacobian of $\vect{H}(\vect{w})$ as input which is computed using the automatic 
differentiation tool ADOL-C \cite{adolc,GJU96}, and described in more detail
in \cite{SBLP13}.

\section{Determining the basin size of the true solution}   \label{sec:determining_basin_size}
%
In Sec. \ref{sec:estimalg} we described a state and parameter estimation 
algorithm that has to be initialized with guesses for all model state
variables $\vect{x}(t_n)$ at each time step $t_n$ and all fixed model parameters 
$\vect{p}$. This set of values forms an \textit{initial guess}, which must be 
supplied to the optimization algorithm. In this 
section three different methods for generating the initial guesses are presented 
and simulations consisting of twin experiments are performed to determine 
which of these methods gives the best estimates for the model state variables and 
fixed parameters. 
These estimates are then compared with the \textit{true solution} which is 
known exactly in this case since this is a twin experiment.
Due to the fact that 
the methods for generating the initial guesses, in a certain way, depend on 
random numbers and the outcome of an estimation process is either successful 
(estimated states and parameters are close to the ones used to generate the data 
time series) or not successful (estimated states and parameters are \textit{not} 
close to the ones used to generate the data time series) the simulations can be 
considered as \textit{Bernoulli experiments} and the basin size of initial 
guesses leading to the true solution can be determined, as suggested by Menck et 
al. \cite{MHMK13} in another context.

\subsection{The simulation} \label{sec:l96_state_par_estimation_simulation}

First we  generate 18 ``true'' trajectories $\{^i\vect{z}(n) \}$ with $i=1, 
\dots, 18$ 
by integrating the 9-dimensional Lorenz-96 model \eqref{eq:lorenz96_model} with 
18 different initial  conditions $\vect{z}(0)$ on the attractor with $\Delta t = 
0.01$ and $N=1500$
using the model parameter $p=8.17$. 
Then $N_\mathrm{iguess} = 500$ initial guesses $(\{{}_h^i\vect{x}(n)\}, {}_h^{}p )$ 
($h = 1, \dots, N_\mathrm{iguess}$) of the model state variables and the (fixed) 
parameter $p$ are 
generated which are used for initializing the  estimation procedure 
(the estimation algorithm was described in Sec. \ref{sec:estimalg}). 
Three different methods for generating the initial guesses will be presented in 
Sec.  \ref{sec:l96_state_par_estimation_initial_guesses}. 
The following steps in the simulation do not depend on the specific 
choice of the method for creating the initial guesses.

From each of the 18 true trajectories $\{^i\vect{z}(n) \}$ with $i=1, \dots, 
18$, according to 
Sec.~\ref{sec:l96_example}, 15 multivariate time series were extracted 
corresponding to the 15 different combinations of state variables assumed 
to measured. This gives 270 different multivariate time series with one, two, 
or three state variables:            
\begin{align}                          
    & \{ {}^iz_1(n) \},
    \{ {}^i(z_1(n),z_2(n)) \}, 
    \{ {}^i(z_1(n),z_3(n)) \}, \notag \\
    & \{ {}^i(z_1(n),z_4(n)) \}, 
    \{ {}^i(z_1(n),z_5(n)) \}, \notag \\                      
    & \{ {}^i(z_1(n),z_2(n),z_3(n)) \}, 
    \{ {}^i(z_1(n),z_2(n),z_4(n)) \}, \notag \\
    &\{ {}^i(z_1(n),z_2(n),z_5(n)) \}, 
    \{ {}^i(z_1(n),z_2(n),z_6(n)) \}, \notag \\
    & \{ {}^i(z_1(n),z_2(n),z_7(n)) \}, 
    \{ {}^i(z_1(n),z_2(n),z_8(n)) \}, \notag \\
    & \{ {}^i(z_1(n),z_3(n),z_5(n)) \}, 
    \{ {}^i(z_1(n),z_3(n),z_6(n)) \}, \notag \\
    & \{ {}^i(z_1(n),z_3(n),z_7(n)) \}, 
    \{ {}^i(z_1(n),z_4(n),z_7(n)) \}
    \label{eq:yset}
\end{align}
with $i=1,\dots, 18$, $\Delta t = 0.01$, $n=0,1, \dots, N$ and 
$N=1500$.

To make the simulation more realistic, white noise (normally distributed random numbers)
is added to these 270 clean, multivariate times series. This results in 270 
noisy multivariate time series $\{ {}^i\vect{\eta}^c(n) \}$
with $\Delta t = 0.01$, $n=0,1, \dots, N$ and $N=1500$. 

Each  noisy time series is computed by
\begin{align}
    {}^i\vect{\eta}^c(n) &= \vect{h}^c\left( {}^i\vect{z}(n) \right) + 
                    \sigma_\mathrm{ts} {}^i\vect{\xi}^c(n) \, ,
    \label{eq:eta_noise}
\end{align}
where $\sigma_\mathrm{ts} = 0.2$ and ${}^i\vect{\xi}^c(n) = \left( 
{}^i\vect{\xi}^c_1(n), \dots, {}^i\vect{\xi}^c_L(n) \right) \in 
\mathbb{R}^L$ are independent, normally distributed random variables with zero 
mean and a variance of one, ${}^i\vect{\xi}^c_l(n) \sim \mathcal{N}(0, 1)$.

The index $i = 1, \dots, 18$ describes the true trajectory 
$\{{}^i\vect{z}(n) \}$ from which the data time series was extracted. Index $c$ 
indicates  which state variables were 
measured. For example, $c=\mathrm{(1-2-6)}$ means that the state variables $z_1$, 
$z_2$ and $z_6$ are the measured state variables. The label $h = 1, \dots, 
N_\mathrm{iguess}$ describes with which initial guess the estimation 
algorithm  was initialized.
The measurement function $\vect{h}^c(\vect{x}(t))$ is always chosen 
according to the measured state variables defined by $c$. If, for example, the 
state variables $z_1,z_2$ and $z_7$ are measured (and therefore $c=1-2-7$), then 
the measurement function is given by $\vect{h}^c(\vect{x}(t))  = (x_1(t),x_2(t),x_7(t))$.

To each of the 270 multivariate time series $\{ {}^i\vect{\eta}^c(n) \}$ the 
Lorenz-96 model is adapted 
$N_\mathrm{iguess}=500$ times, whereas each of the $N_\mathrm{iguess}$ 
estimation processes is initialized with one of the previously generated 
$N_\mathrm{iguess}$ different (random) initial guesses using the estimation 
algorithms described in Sec.~\ref{sec:estimalg}. This means that 
$N_\mathrm{iguess} \cdot 270 = 500 \cdot 270 = 135000$ estimation problems 
are solved. 

For each solution of the estimation processes the difference between the true 
and the estimated solution is given by the \textit{estimation error}
\begin{align}
    {}_h^iE^c =& \frac{1}{(N+1)\cdot D} \sum_{n=0}^N
            \left\Vert {}^i\vect{z}(n) 
            - {}_h^i\hat{\vect{x}}^c(n) \right\Vert^2_2 \, .
    \label{eq:A}
\end{align}
The indices of ${}_h^iE^c$, ${}^i\vect{z}^c(n)$ and ${}_h^i\hat{\vect{x}}^c(n)$ 
have the same 
meaning as for $\{ {}_h^i\hat{\vect{x}}^c(n) \}$. The smaller the error measure 
${}_h^iE^c$ the 
closer the estimated solution for the model state variables is to the true solution 
and hence the more accurately the estimation problem was solved.  The 
estimation of state variables is considered as successful if ${}_h^iE^c < 10^{-2}$,  
else the estimation is considered as \textit{not} successful.  
The value for the estimated fixed model parameter ${}_h^i\hat{p}^c$ is considered as 
successful, if 
${}_h^i\hat{p}^c \in [8.16,8.18]$ (remember, the true trajectories $\{ 
{}^i\hat{\vect{z}}(n) \}$
were generated with $p=8.17$ in Eq.~\eqref{eq:lorenz96_model}). 

We are interested in a quantity (in percentage) which tells us how many 
estimations 
with a specific true trajectory, $i$, and a specific combination of observed state
variables, $c$, of the model state variables are successful. In other words: For how
many of the $N_\mathrm{iguess} = 500$ estimations using $N_\mathrm{iguess}$ 
different initial guesses with a specific true trajectory, $i$, and a 
specific combination of observed state variables, $c$, is the estimation of the model state
variables successful, i.e. ${}_h^iE^c < 10^{-2}$? This quantity, which of course
depends on the true trajectory and the combination of measured state variables, is 
defined here as the \textit{success rate of the estimation} of the model state
variables (in percentage)
\begin{align}
    \label{eq:A_average}
    \left< {}^iE^c \right> &= \frac{100 \%}{N_\mathrm{iguess}} 
                                \sum_{h=1}^{N_\mathrm{iguess}} e_h, \\
                \mathrm{with} \quad e_h &= 
    \left\{\begin{matrix}
                0, & \mathrm{if} \; {}_h^iE^c > 10^{-2} \\
                1, & \mathrm{if} \; {}_h^iE^c \leq  10^{-2}
            \end{matrix}
    \right. \; .
    \notag
\end{align}
One can also define an error which
depends on the estimated solution and the data, only, as
\begin{align}
    {}_h^iE^c_\mathrm{obs} =& \frac{1}{(N+1)\cdot L} \sum_{n=0}^N
            \left\Vert {}^i\vect{\eta}^c(n) 
            - \vect{h}^c\left({}_h^i\hat{\vect{x}}^c(n)\right) \right\Vert^2_2 
            \; . 
    \label{eq:Eobs}
\end{align}
Assume one estimates the best possible solution. That is, if the estimated 
solution is equal to the trajectory (without noise) used to generate the data, 
${}_h^i\hat{\vect{x}}^c(n) = {}^i\vect{z}(n)$. In this case 
Eq. \eqref{eq:Eobs} is (using eq. \eqref{eq:eta_noise})
\begin{align}
    {}_h^iE^c_\mathrm{obs,opt} =& \frac{1}{(N+1) L} \sum_{n=0}^N
            \left\Vert {}^i\vect{\eta}^c(n) 
            - \vect{h}^c\left({}^i\vect{z}^c(n)\right) \right\Vert^2_2 \notag \\
            =& \frac{\sigma_\mathrm{ts}^2}{(N+1) L} 
                {}^iQ^c \; ,
    \label{eq:Eobsopt}
\end{align}
where
\begin{align}
{}^iQ^c =& \sum_{n=0}^N \sum_{l=1}^L \left[ 
                {}^i\xi^c_l(n) \right]^2 \; . 
\end{align}
Because ${}^i\xi^c_l(n)$ are  independent, standard 
normal random variables, ${}^iQ^c$ is chi-squared distributed, 
${}^iQ^c \sim \chi^2_{(N+1)L}$, with $(N+1)L$ degrees of 
freedom\cite{numerical_recipes_2007}. The 
expectation value is then given by $E \left[ {}^iQ^c \right] = (N+1)L$, leading 
to an expectation value for ${}_h^iE^c_\mathrm{obs,opt}$ of
\begin{align}
    E \left[ {}_h^iE^c_\mathrm{obs,opt} \right] &= \sigma_\mathrm{ts}^2 \; .
\end{align}
The variance of the chi-square distribution is $\var \left[ {}^iQ^c \right]  = 
2(N+1)L $. The variance of ${}_h^iE^c_\mathrm{obs,opt}$ is 
then
\begin{align}
    \var \left[ {}_h^iE^c_\mathrm{obs,opt} \right] 
    &= \var \left[ \frac{\sigma_\mathrm{ts}^2}{(N+1) L} 
        {}^iQ^c \right] \notag \\
    &= \frac{\sigma_\mathrm{ts}^4}{[(N+1) L]^2}\var \left[  
        {}^iQ^c \right] \notag \\
    &= \frac{2\sigma_\mathrm{ts}^4}{(N+1) L}
\end{align}
giving the standard deviation
\begin{align}
    \std \left[ {}_h^iE^c_\mathrm{obs,opt} \right] 
    &= \sqrt{\var \left[ {}_h^iE^c_\mathrm{obs,opt} \right]} \notag \\
    &= \sigma_\mathrm{ts}^2 \sqrt{\frac{2}{(N+1) L}} \; .
\end{align}
As described in Sec. \ref{sec:l96_state_par_estimation_simulation} we use 
$\sigma_\mathrm{ts} = 0.2$ and $N=1500$. For one observed state variable, $L=1$, and 
a perfect solution of the estimation problem, we get the lower boundary for 
${}_h^iE^c_\mathrm{obs}$, Eq. \eqref{eq:Eobs}, of
\begin{align}
    \sigma_\mathrm{ts}^2 \left[1 \pm  \sqrt{\frac{2}{(N+1)L)}} \right] 
    \approx 0.04 \pm 0.00146 \; .
\end{align}
Note that only the standard deviation depends on the number of measurements 
(and is largest for $L=1$), but not the expectation value.
This means that with a smooth estimate for the model state variables one
can not go below this boundary. If one goes below this threshold, the 
measurement noise is modelled and one has not estimated a smooth solution for 
the model variables. In this case one should choose a smaller $\alpha$ in the 
cost function Eq. \eqref{eq:cost}. Note that the modelling of the measurement 
noise is still possible if one does not fall below this boundary. Because of 
the perfect model scenario in our twin experiments we can expect a value for 
${}_h^iE^c_\mathrm{obs}$ which is only slightly larger (due to small numerical 
errors) than the lower boundary of $ 0.04 \pm 1.46 \cdot 10^{-3}$. 
To cover these cases we introduce an empirical margin of 0.005 
which is added to the lower bound of 0.04 and we
consider an estimation as successful if ${}_h^iE^c_\mathrm{obs} \leq 0.045$. 
Applying this bound to the error given by Eq.~\eqref{eq:Eobs} we can define a 
success rate (in percentage)
\begin{align}
    \label{eq:Eobs_average}
    \left< {}^iE_\mathrm{obs}^c \right> &= \frac{100 \%}{N_\mathrm{iguess}} 
                                \sum_{h=1}^{N_\mathrm{iguess}} e_h, \\
                \mathrm{with} \quad e_h &= 
    \left\{ \begin{matrix}
                0, & \mathrm{if} \; {}_h^iE_\mathrm{obs}^c > 0.045 \\
                1, & \mathrm{if} \; {}_h^iE_\mathrm{obs}^c \leq  0.045
            \end{matrix}
    \right. \; .
    \notag
\end{align}

In a similar way, the \textit{success rate of the estimation of the model 
parameter} (in percentage) ${}_h^i\hat{p}^c$ is defined as
\begin{align}
    \label{eq:p_average}
    \left< {}^i\hat{p}^c \right> &= \frac{100 \%}{N_\mathrm{iguess}} 
                                \sum_{h=1}^{N_\mathrm{iguess}} p_h, \\
    \mathrm{with} \quad p_h &= 
    \left\{\begin{matrix}
                0, & \mathrm{if} \; {}_h^i\hat{p}^c \notin [8.16,8.18] \\
                1, & \mathrm{if} \; {}_h^i\hat{p}^c \in [8.16,8.18]
            \end{matrix}
    \right. \; .
    \notag
\end{align}
Note, that in a real world experiment ${}_h^iE^c$, and hence $\left< {}^iE^c 
\right>$,  typically can \textit{not} be computed due to the unknown true trajectory $\{ {}^i\vect{z}(n) \}$. 
Another possibility to compute the accuracy of the estimated model state variables and the fixed
model parameters is to compare predictions of the model via the measurement 
function $\vect{h}(\vect{x}(t))$, Eq.~\eqref{eq:measurementfct}, with available 
(noisy) data after the estimation window. To compute the prediction the model 
Eq.~\eqref{eq:ode_model} must be integrated starting at the end of the 
estimation window at $t_N$ using the estimated value ${}_h^i\hat{p}^c$ as model 
parameter and ${}_h^i\hat{\vect{x}}^c(N)$ as initial guesses. Next, the 
prediction $\{ {}_h^i\vect{x}^c(n) \}$, $n \geq N$, can be compared with 
observed data $\{ {}^i\vect{\eta}^c(n) \}$, $n \geq N$, by computing the 
\textit{prediction error} 
\begin{align}
    {}_h^i\mathrm{PE}^c =& \frac{1}{(N_\mathrm{pred}+1)\cdot L} 
                            \sum_{n=N}^{N + N_\mathrm{pred}}
                            \left\Vert {}^i\vect{\eta}^c(n) 
                            - \vect{h}^c \left( {}_h^i\vect{x}^c(n) \right) 
                            \right\Vert^2_2
    \label{eq:PE}
\end{align}
for $N_\mathrm{pred}$ time steps using the same step size $\Delta t$ as for 
computing the true trajectories. Due to noise in the data the prediction error cannot vanish and  
we consider a prediction as successful, if  ${}_h^i\mathrm{PE}^c < 0.5$. 
Analogous to Eq.~\eqref{eq:A_average} we define the \textit{success rate of the 
prediction} (in 
percentage)
\begin{align}
    \label{eq:PE_average}
    \left< {}^i\mathrm{PE}^c \right> &= \frac{100 \%}{N_\mathrm{iguess}} 
                \sum_{h=1}^{N_\mathrm{iguess}} \mathrm{pe}_h, \\
    \mathrm{with} \quad \mathrm{pe}_h &= 
    \left\{\begin{matrix}
                0, & \mathrm{if} \; {}_h^i\mathrm{PE}^c > 0.5 \\
                1, & \mathrm{if} \; {}_h^i\mathrm{PE}^c \leq 0.5
            \end{matrix}
    \right.
    \notag
\end{align}
describing for how many of the different initial guesses with 
the same true trajectory and the same combination of measured state variables the 
prediction was successful.
In contrast to ${}_h^iE^c$ the prediction 
error ${}_h^i\mathrm{PE}^c$ can be computed using measured data only.

\subsection{Different methods for generating  initial guesses}
%
\label{sec:l96_state_par_estimation_initial_guesses}
For the optimization process initial guesses  for the model state variables and 
the fixed model parameter $p$ have to be chosen. In our simulation we considered 
three methods for preparing initial guesses according to rules specified below.
For each case $N_\mathrm{iguess} = 500$ different guesses
$(\{{}_h^i\vect{x}(n)\}, {}_h^{}p )$ with $h = 1, \dots, N_\mathrm{iguess}$ 
are generated. In all three cases the model 
parameter ${}_hp$ is picked equally distributed from the interval $[4,20]$. 
In those cases where the initial guess $\{{}_h^i\vect{x}(n)\}$ for the 
model state variables 
does \textit{not} depend on the true trajectory $\{{}^i\vect{z}(n)\}$ of the 
estimation problem, the index $i$ will be neglected (i.e. 
${}_h^i\vect{x}(n) = {}_h\vect{x}(n)$). In the following, three different 
methods of choosing the initial guesses will be used and evaluated:
\begin{enumerate}
    \item   \textbf{Uniformly distributed samples in a box:}
            \label{enum:iguess1}
            For each initial guess each model state variable ${}_h^{}x_d(n)$,
            $d=1, \dots, D$ at each time step $t_n$ is an equally distributed 
            random number in the interval $[-9,14]$. This interval has been 
            chosen because it is the range of typical oscillations of all 
            state variables of the Lorenz-96 model.
            Together with the model parameter the initial guesses consist of 
            $D \cdot N+1=D_\mathrm{iguess}=13501$ numerical values.
            In other words, the initial guesses are uniformly distributed 
            points in a box in a $D_\mathrm{iguess}$ dimensional space           
            $\mathbb{R}^{D_\mathrm{iguess}}$.            
    \item   \textbf{Exact solutions of the model:}
            \label{enum:iguess2}
            Each initial guess $(\{{}_h^{}\vect{x}(n)\}, {}_h^{}p )$ is an 
            exact solution of the Lorenz-96 model Eq.~\eqref{eq:lorenz96_model}.
            The initial values ${}_h^{}\vect{x}(0)$ of these trajectories are arbitrary points on the 
            attractor generated with $p=8.17$ (not coinciding with the initial 
            conditions of the true trajectories).
            
    \item   \textbf{Samples close to the true solution:}
            \label{enum:iguess3}
            These initial guesses depend, in contrast to methods \ref{enum:iguess1}
            and \ref{enum:iguess2}, on the ``true trajectories'' 
            $\{{}^i\vect{z}(n)\}$ with $i=1, \dots, 18$ (see 
            Sec.~\ref{sec:l96_state_par_estimation_simulation}). 
            The estimation processes will be initialized with a ``noisy'' 
            version of $\{{}^i\vect{z}(n)\}$. More precisely, for each time step 
            $t_n$ uniformly distributed random numbers 
            from the interval $[ -15, 15]$ are added to the values of the true state
            $\{{}^i\vect{z}(n)\}$ to generate the initial guesses 
            $\{{}_h^i\vect{x}(n)\}$,
            Compared to initial guess strategy \ref{enum:iguess1} and 
            \ref{enum:iguess2} this strategy does depend on the true 
            trajectories. In a real world application, where the true 
            trajectories are not known, this strategy can not 
            be used in contrast to methods \ref{enum:iguess1} and                 
            \ref{enum:iguess2}.
\end{enumerate}

\subsection{Interpretation of the simulation as Bernoulli experiment and error estimation}
%
As described in Sec.~\ref{sec:l96_state_par_estimation_simulation} 
for each of the 18 true trajectories
and each of the 15 combinations of measured state variables, the Lorenz-96 model was 
adapted $N_\mathrm{iguess}=500$ times to the corresponding (multivariate) time 
series using a specific method for choosing the initial guesses. If 
${}_h^iE^c < 10^{-2}$ (Eq.~\eqref{eq:A}) then the estimation of the model state
variables is considered as successful. This simulation can be interpreted as  a 
Bernoulli experiment, because each of the independent 
$N_\mathrm{iguess}$ estimations of the model state variables 
and the fixed parameter is a Bernoulli trial with the outcome \textit{successful} or 
\textit{not successful}. The standard error of the Bernoulli process is given by
\begin{align}
    {}^ie^c &:= \frac{\sqrt{{}^ip^c(100\%-{}^ip^c)}}
                {\sqrt{N_\mathrm{iguess}}} \, ,
    \label{eq:bernoulli_stderr}
\end{align}
whereas ${}^ip^c \in [0\%, 100\%]$ is the expectation value of the percentage of  
successful cases (index $i$ describes the used true trajectory and 
index $c$ describes the combination of measured state variables). Unfortunately, we 
do not know  ${}^ip^c$. 
However, we can determine the maximum of the standard error ${}^ie^c$ which occurs 
for ${}^ip^c = 50\%$. With $N_\mathrm{iguess}=500$ trials the maximal standard error equals
${}^ie^c_\mathrm{max} \approx 2.24\%$ and hence is sufficiently small.

\subsection{Results}    \label{sec:l96_state_par_estimation_results}
%

\subsubsection{Estimation Error}

The simulation described in Sec.~\ref{sec:l96_state_par_estimation_simulation} 
was performed with all three methods for choosing initial guesses for the 
model state variables and the fixed model parameters $(\{{}_h^i\vect{x}(n)\}, {}_h^{}p)$ 
as described in Sec.~\ref{sec:l96_state_par_estimation_initial_guesses}. 
For each method of choosing the initial guesses the percentage of successful 
estimations, $\left< {}^iE^c \right>$, Eq.~\eqref{eq:A_average},  was computed, 
where an estimation of the model state variables is 
considered as successful if ${}_h^iE^c <  10^{-2}$, 
Eq.~\eqref{eq:A} (see Sec.~\ref{sec:l96_state_par_estimation_simulation}). 
The estimation of the model parameter is considered as successful if  
${}_h^i\hat{p}^c \in [8.16,8.18]$. The success rate for the fixed model parameter, 
$\left< {}^i\hat{p}^c \right>$, is defined in Eq.~\eqref{eq:p_average}.
The statistic (percentage of successful 
estimations) was created for each of the 18 true trajectories $\{^i\vect{z}(n) 
\}$ (indexed by $i$), each of the 15 combinations of observed state
variables, $c$, and all $N_\mathrm{iguess} = 500$ initial guesses 
(indexed by $h$). 

Tables~\ref{tab:z_N1800_estim_succsessful}a,b show the results for 
method \ref{enum:iguess1} (uniformly distributed samples in a box). 
The tables show $\left< {}^iE^c \right>$ 
(Table~\ref{subtab:variables_observed}) and $\left< {}^i\hat{p}^c 
\right>$ (Table~\ref{subtab:parameter_observed}) for each  combination 
of a true trajectory $\{^i\vect{z}(n) \}$ and a particular choice of measured 
state variables. If three variables are measured, 
the rate of successful estimations of the model variables and the fixed parameter is 
(on average)  higher for all combinations of measured state variables compared to the success rate for
multivariate time series with only two variables. 
%
%
\begin{figure*} 
    \centering
    \includegraphics[width=1\textwidth]{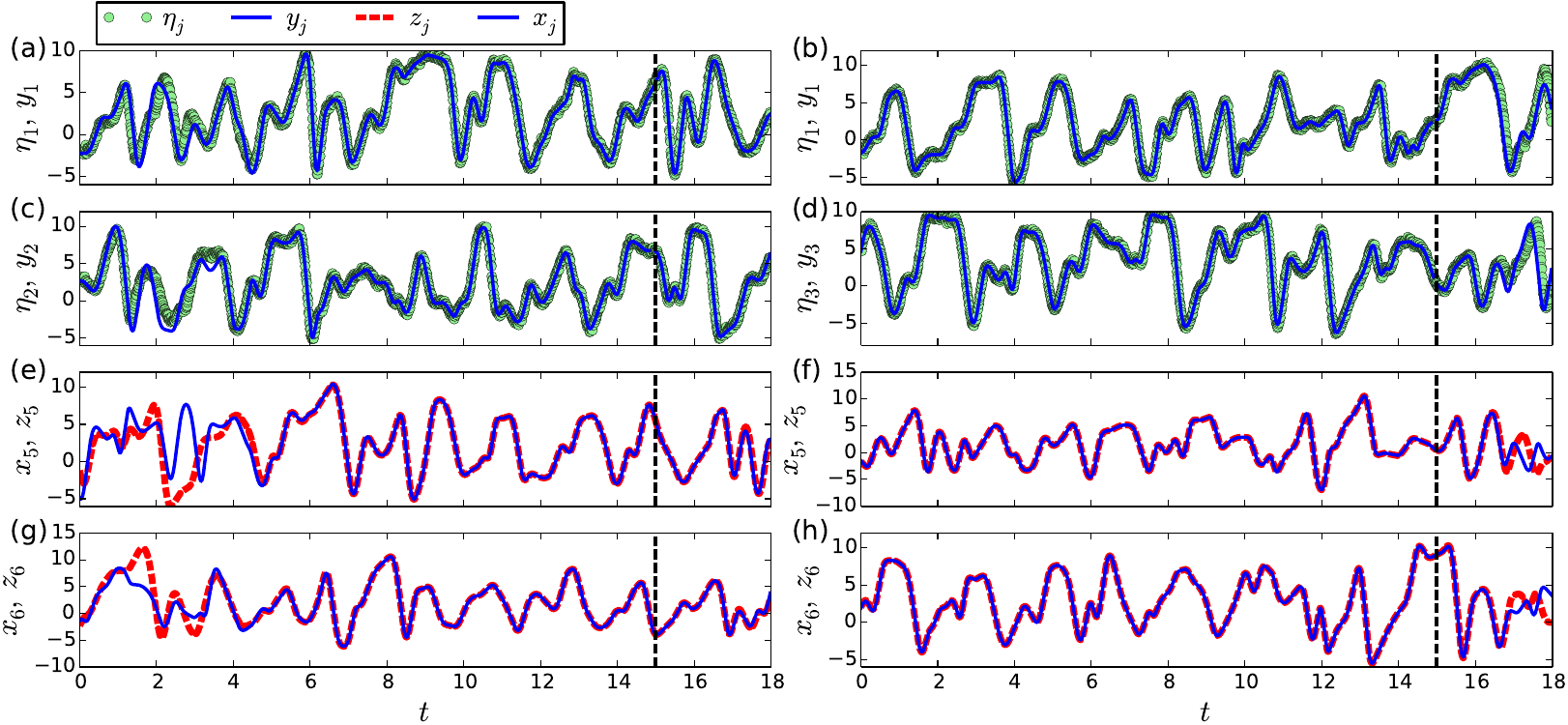}
    \caption{This figure show two examples where the Lorenz-96 model 
            Eq.~\eqref{eq:lorenz96_model} was adapted to a (multivariate) time 
             series $\eta_j$. The (unmeasured) model state variables $x_j$ and the 
            fixed model parameter were estimated using the estimation 
            method 1 described
            in Sec.~\ref{sec:estimalg}. The output of the measurement function 
            is $y_j$ and the true trajectory is $z_j$ (unknown to the 
            estimation algorithm).  The estimation was performed for
            $0 < t < 15$ and the prediction of the model variables for $15 \leq 
            t < 18$ (right of the vertical black dashed line at $t=15$).
            Left column, (a), (c), (e), (g): $x_1$ and $x_2$ are measured
            ($c=\mathrm{(1-2)}$) and $i=3$, $h=385$. The estimation error of 
            model variables is larger than the prediction error,
            ${}_{385}^3E^\mathrm{(1-2)} = 1.73 > 
            {}_{385}^3\mathrm{PE}^\mathrm{(1-2)} = 0.054$.
            Right column, (b), (d), (f), (h): $x_1$ and $x_3$ are measured
            ($c=\mathrm{(1-3)}$) and $i=1$, $h=140$. The estimation error of 
            model variables is smaller than the prediction error,
            ${}_{140}^1E^\mathrm{(1-3)} = 6.1 \cdot 10^{-4} < 
            {}_{140}^1\mathrm{PE}^\mathrm{(1-3)} = 1.98$.}
    \label{fig:prediction_error}
\end{figure*}
Nevertheless, certain combinations with two observed variables $(x_1,x_2)$, 
$(x_1,x_3)$ or $(x_1,x_4)$ also give success rates that are only slightly lower 
than combinations of three observed state variables.
They just appear less often compared to time series with three observed state 
variables. 
When $(x_1,x_5)$ are observed the estimation of the model state variables and the 
fixed parameter does not seem to work very well. 
None of the 18 trajectories considered here exhibit high success rate.
If only $x_1$ is observed the estimation of variables and 
the parameter fails for all 18 trajectories. As one might expect, one can see 
a high correlation between the success rate for the state variable estimation 
(Tab.~\ref{subtab:variables_observed})  and the 
success rate for the parameter estimation  (Tab.~\ref{subtab:parameter_observed}). 
The success rates depend not only on  
the combination of observed variables only, but also on the 
trajectory $\{ ^i\vect{z}(n) \}$ used to generate the time series (i.e. 
the starting points on the attractor). 

In Tab. \ref{tab:variables_observed_estim_error} the success rate of the error 
defined by Eq.~\eqref{eq:Eobs_average} is shown. Compared to 
Eq.~\eqref{eq:A_average} this success rate can be computed from the data and 
the estimated model state variables only. One can see a high correlation between 
Tab.~\ref{subtab:variables_observed} and 
Tab.~\ref{tab:variables_observed_estim_error} indicating that 
${}_h^iE^c_\mathrm{obs}$ is a good approximation of ${}_h^iE^c$ (at least in 
the absence of errors in the model equations, as in these simulations). There 
are, however, some discrepancies. For example, if $\{ {}^5\vect{z}(n) \}$ is 
the true trajectory and $c = 1 - 2$ is measured, 
Tab.~\ref{subtab:variables_observed} shows a much smaller success rate, 
given by Eq.~\eqref{eq:A_average} (only the error of all model variables is 
considered), compared to the success rate, given by 
Eq.~\eqref{eq:Eobs_average}, in Tab.~\ref{tab:variables_observed_estim_error} 
(the error of measured state variables is considered only). This shows that a good 
estimation of measured variables does not necessarily mean that unmeasured 
variables are also estimated correctly. 

With initial guess method \ref{enum:iguess1}  the initial guess for 
each of the 9 model variable at 1500 locations along the (initial) trajectory 
is a random number (equally distributed) from the interval $[-9,14]$ 
(see Sec.~\ref{sec:l96_state_par_estimation_initial_guesses}). With the guess 
for the unknown parameter the full initial guess  is a point in a 
$D_\mathrm{iguess}=13501$ dimensional box.  
Scanning this entire $D_\mathrm{iguess}=13501$ dimensional rectangular box 
containing the initial is not an appropriate method to learn something about the basin shape
of the optimal solution. Nevertheless, 
one can interpret the success rate $\left< {}^iE^c \right>$ as the ratio of the 
size of the basin of successful estimates and the volume of the box in the 
$\mathbb{R}^{D_\mathrm{iguess}}$ space \cite{MHMK13} in percentage.
\begin{table*}\footnotesize
    \centering
    \subfloat[][Observability of model state variables]{
    \label{subtab:variables_observed}
    \begin{tabular}{c*{15}{r}}
         \toprule    
         \textbf{True }& \multicolumn{15}{c}{\textbf{Observed state variables $c$}}\\ 
         \textbf{trajectory} & \text{1} & \text{1-2} &  
         \text{1-3} & \text{1-4} & \text{1-5}
         & \text{1-2-3} & \text{1-2-4} & \text{1-2-5} & \text{1-2-6} 
         & \text{1-2-7} & \text{1-2-8} & \text{1-3-5} & \text{1-3-6}
         & \text{1-3-7} & \text{1-4-7} \\       
        \colrule
$\{ {}^{1}\vect{z}(n) \}$ &     0&  49.8&   33&     42.6&   0&  95.6&   98.8&   
86.8&   95.8&   83.6&   8.6&    50.2&   85.4&   90.2&   89.8 \\
$\{ {}^{2}\vect{z}(n) \}$ &     0&  3.6&    0&  1.6&    0&  2.6&    89.8&   
90.2&   18.4&   65&     17.4&   8.6&    5.2&    15.6&   64.8 \\
$\{ {}^{3}\vect{z}(n) \}$ &     0&  1.8&    5.4&    0&  0.2&    66.2&   82.2&   
35&     2&  39.6&   3.4&    17.6&   76.4&   87.8&   83.4 \\
$\{ {}^{4}\vect{z}(n) \}$ &     0&  1&  0.4&    0.2&    0&  77.8&   94.2&   
90.2&   2&  87.8&   34.6&   34.8&   31&     28.8&   18.2 \\
$\{ {}^{5}\vect{z}(n) \}$ &     0&  8.4&    60.6&   0.6&    5.2&    74.8&   95&  
   90&     83.2&   93&     41&     86.6&   84.6&   88.2&   89.8 \\
$\{ {}^{6}\vect{z}(n) \}$ &     0&  37.4&   0.2&    0&  0.2&    58.2&   77.6&   
80.8&   94.6&   56.2&   78.8&   7.8&    27.8&   14.4&   25.4 \\
$\{ {}^{7}\vect{z}(n) \}$ &     0&  2.8&    2.2&    6.4&    0&  96.4&   93.6&   
76.4&   91.4&   30&     30.2&   37.8&   55.2&   36.4&   82.2 \\
$\{ {}^{8}\vect{z}(n) \}$ &     0&  95.2&   5.2&    0.8&    1&  97.2&   98.2&   
92.4&   85.6&   86.6&   94.4&   83.2&   65.2&   59.8&   89.2 \\
$\{ {}^{9}\vect{z}(n) \}$ &     0.2&    87.8&   80.2&   76&     0.2&    96.8&   
99.4&   87.4&   89&     87.8&   12.4&   89.4&   83.4&   77.8&   89.2 \\
$\{ {}^{10}\vect{z}(n) \}$ &    0&  92.4&   28.4&   1.2&    2.8&    98.2&   
98.4&   94.8&   85.4&   50.4&   81.8&   81&     68.4&   6.2&    4.2 \\
$\{ {}^{11}\vect{z}(n) \}$ &    0&  0&  0.4&    0.6&    5.2&    4&  63.8&   
89.2&   1.6&    60.8&   61.8&   58.8&   49.4&   81&     87.2 \\
$\{ {}^{12}\vect{z}(n) \}$ &    0&  24.8&   1.2&    1.6&    0&  88.6&   95.4&   
89&     1.6&    3.6&    12.2&   20.8&   2.6&    0.6&    86.4 \\
$\{ {}^{13}\vect{z}(n) \}$ &    0&  4.8&    27.2&   0.2&    3.6&    82.8&   
95.6&   81&     7.2&    92.6&   86&     67.2&   67.8&   66.4&   86 \\
$\{ {}^{14}\vect{z}(n) \}$ &    0&  30.6&   7.8&    1.4&    0&  90.6&   95.8&   
78&     83.8&   6.6&    79&     56.2&   68&     42.6&   88.6 \\
$\{ {}^{15}\vect{z}(n) \}$ &    0&  47.2&   1&  0.2&    0&  85&     96.4&   
89.4&   44.6&   85.8&   87.4&   66.8&   36.4&   58.2&   82.2 \\
$\{ {}^{16}\vect{z}(n) \}$ &    0&  14.4&   14.2&   0&  0&  95.4&   96&     94&  
   59.4&   76.6&   96.6&   27.6&   9&  15.6&   7.2 \\
$\{ {}^{17}\vect{z}(n) \}$ &    0&  78.4&   1.4&    0.4&    0.2&    83.6&   
92.4&   95&     92.4&   5.4&    28.4&   16.6&   3.4&    1.2&    87.8 \\
$\{ {}^{18}\vect{z}(n) \}$ &    0&  37.6&   2.2&    1.6&    0&  87.6&   96.6&   
86&     86.2&   71&     32.8&   16.4&   22.8&   71.8&   86.2 \\
    \botrule
    \end{tabular}
    }
    \vspace{0.7cm}    
    \subfloat[][Observability of the fixed model parameter]{
    \label{subtab:parameter_observed}
    \begin{tabular}{c*{15}{r}}        
        \toprule 
         \textbf{True} & \multicolumn{15}{c}{\textbf{Observed state variables $c$}}\\ 
         \textbf{trajectory} & \text{1} & \text{1-2} &  \text{1-3} & 
         \text{1-4} & \text{1-5} & \text{1-2-3} & \text{1-2-4} & 
         \text{1-2-5} & \text{1-2-6} & \text{1-2-7} & \text{1-2-8} & 
         \text{1-3-5} & \text{1-3-6} & \text{1-3-7} & \text{1-4-7} \\   
        \colrule
$\{ {}^{1}\vect{z}(n) \}$ &     0&  54.2&   33&     46.2&   0&  95.8&   99&     
87&     95.8&   83.6&   33.8&   51.2&   85.4&   90.2&   89.8 \\
$\{ {}^{2}\vect{z}(n) \}$ &     0&  4.2&    0&  6&  0.6&    32.2&   89.8&   
90.2&   18.4&   65&     17.6&   8.8&    5.2&    16&     65.4 \\
$\{ {}^{3}\vect{z}(n) \}$ &     0.2&    3.6&    6&  1.4&    3.4&    66.6&   
93.6&   89.6&   2&  39.6&   7.6&    20.6&   81.4&   87.8&   90 \\
$\{ {}^{4}\vect{z}(n) \}$ &     0.2&    1.4&    0.6&    0.4&    0.8&    77.8&   
94.2&   90.2&   2&  88.4&   46.4&   35.2&   31&     30.6&   18.4 \\
$\{ {}^{5}\vect{z}(n) \}$ &     0.6&    67.2&   61.8&   10.4&   5.6&    90.2&   
95.2&   90&     83.2&   93&     41.4&   86.6&   85.2&   92.4&   89.8 \\
$\{ {}^{6}\vect{z}(n) \}$ &     0.2&    37.4&   0.4&    0.2&    0.2&    58.2&   
77.6&   80.8&   94.6&   56.2&   78.8&   7.8&    28.2&   15&     25.4 \\
$\{ {}^{7}\vect{z}(n) \}$ &     0.2&    2.8&    26&     6.6&    3.8&    96.4&   
93.8&   77&     92&     32.8&   56&     40&     55.2&   36.8&   82.6 \\
$\{ {}^{8}\vect{z}(n) \}$ &     0.4&    95.2&   6.2&    1.2&    1.8&    97.6&   
98.4&   92.4&   85.6&   86.8&   95&     83.8&   69.2&   61.6&   94.4 \\
$\{ {}^{9}\vect{z}(n) \}$ &     0.8&    91&     81.2&   79.8&   0.2&    97&     
99.4&   88.2&   93.8&   88&     55.2&   89.8&   83.8&   90.4&   89.4 \\
$\{ {}^{10}\vect{z}(n) \}$ &    0&  92.4&   59&     4.2&    18.6&   98.2&   
98.4&   94.8&   87&     50.4&   81.8&   84.6&   68.4&   53&     92.6 \\
$\{ {}^{11}\vect{z}(n) \}$ &    0.6&    1.2&    1.8&    0.6&    14.6&   4.2&    
63.8&   90.4&   23&     61.2&   63.4&   59.2&   57.8&   82&     87.6 \\
$\{ {}^{12}\vect{z}(n) \}$ &    0&  53.8&   1.4&    5.8&    1.2&    88.8&   
95.4&   89&     2.4&    4.2&    14.8&   45.8&   3.8&    0.8&    88 \\
$\{ {}^{13}\vect{z}(n) \}$ &    0.2&    6.6&    27.8&   0.4&    23.2&   82.8&   
96.4&   84&     60.8&   93&     86&     67.2&   75.4&   66.8&   86.2 \\
$\{ {}^{14}\vect{z}(n) \}$ &    1.4&    31.4&   11.6&   2.8&    0.8&    93.8&   
97&     83.4&   83.8&   6.8&    83&     57.6&   68.2&   42.8&   89.4 \\
$\{ {}^{15}\vect{z}(n) \}$ &    0.4&    59&     7&  1.2&    1.2&    98.8&   97&  
   89.8&   44.6&   86.2&   89&     67&     36.8&   58.2&   82.6 \\
$\{ {}^{16}\vect{z}(n) \}$ &    0.8&    15&     14.4&   10.4&   0.8&    95.6&   
96.6&   94&     59.4&   83.2&   97.4&   36&     9.8&    85.6&   9.8 \\
$\{ {}^{17}\vect{z}(n) \}$ &    0.4&    87.8&   2&  0.6&    0.4&    84.4&   
92.4&   95&     92.4&   44&     30&     16.8&   3.4&    43.8&   87.8 \\
$\{ {}^{18}\vect{z}(n) \}$ &    0&  39.6&   2.4&    2&  1.6&    88.4&   96.6&   
86&     86.2&   92.4&   32.8&   16.6&   23.2&   72&     86.8 \\
    \botrule
    \end{tabular}
    }
    \caption{These tables show the results of the simulation explained in 
            Sec. \ref{sec:l96_state_par_estimation_simulation} with initial 
            guess method Sec. 
            \ref{sec:l96_state_par_estimation_initial_guesses} method 
            \ref{enum:iguess1}. 
            For the 9 dimensional
            Lorenz-96 model Eq.~\eqref{eq:lorenz96_model} there exist 
            15 mathematically different combinations of one to three
            state variables constituting a multivariate time series
            (Sec.~\ref{sec:l96_state_par_estimation_simulation}, 
            the first rows of tables (a) and (b) show all these combinations). 
            Example: 1-2-4 means that          
            the variables $z_1$, $z_2$ and $z_4$ are measured.      
            The 18 noise-free time series $\{ {}^i\vect{z}(n) \}$, $i=1, \dots, 18$ 
            are generated by
            integrating the model equations with different initial conditions.  
            For each $i$, from $\{ {}^i\vect{z}(n) \}$ we extract 15 
            different time series with different combinations of state variables. 
            According to Eq.~\eqref{eq:eta_noise}, some 
            artificial noise is added (Sec. \ref{sec:l96_state_par_estimation_simulation}). 
            This results in $15 \cdot 18 = 270$ 
            different noisy multivariate time series (cf. 
            Eq.~\eqref{eq:eta_noise}). 
            To each of the 270 noisy time series the Lorenz-96 model 
            is adapted $N_\mathrm{iguess}=500$ times using the state and 
            parameter estimation algorithm described in Sec. \ref{sec:estimalg} 
            with 500 different initial guesses for the model state variables and the 
            fixed model parameter chosen according to initial guess method 1
            (uniformly distributed samples in a box) 
            (Sec. \ref{sec:l96_state_par_estimation_initial_guesses}).
            For each of the 500 solutions ${}_h^iE^c$ (Eq.~\eqref{eq:A})
            is computed ($h=1,\dots, 500$). If ${}_h^iE^c <  10^{-2}$, 
            then the variables estimation is considered as successful.
            The values in the tables show the percentages of successful 
            estimations of (a) state variables, $\left< {}^iE^c \right>$
            Eq.~\eqref{eq:A_average}, and (b) parameters,
            $\left< {}^i\hat{p}^c \right>$ Eq.~\eqref{eq:p_average}.}     
\label{tab:z_N1800_estim_succsessful}
\end{table*}
\begin{table*}\footnotesize
    \centering   
    \begin{tabular}{c*{15}{r}}
         \toprule    
         \textbf{True }& \multicolumn{15}{c}{\textbf{Observed state variables $c$}}\\ 
         \textbf{trajectory} & \text{1} & \text{1-2} &  
         \text{1-3} & \text{1-4} & \text{1-5}
         & \text{1-2-3} & \text{1-2-4} & \text{1-2-5} & \text{1-2-6} 
         & \text{1-2-7} & \text{1-2-8} & \text{1-3-5} & \text{1-3-6}
         & \text{1-3-7} & \text{1-4-7} \\       
        \colrule
$\{ {}^{1}\vect{z}(n) \}$ &     0&  49.8&   33&     42.6&   0&  95.6&   98.8&   
86.8&   95.8&   83.6&   33&     50.2&   85.4&   90.2&   89.8 \\
$\{ {}^{2}\vect{z}(n) \}$ &     0&  3.6&    0&  1.6&    0&  2.6&    89.8&   
90.2&   18.4&   65&     17.4&   8.6&    5.2&    15.6&   64.8 \\
$\{ {}^{3}\vect{z}(n) \}$ &     0&  1.8&    5.4&    0&  0.2&    66.2&   82.2&   
35&     2&  39.6&   3.4&    17.6&   76.4&   87.8&   83.4 \\
$\{ {}^{4}\vect{z}(n) \}$ &     0&  1&  0.4&    0.2&    0&  77.8&   94.2&   
90.2&   2&  87.8&   34.6&   34.8&   31&     28.8&   18.2 \\
$\{ {}^{5}\vect{z}(n) \}$ &     0&  67&     60.6&   0.6&    5.2&    74.8&   95& 
 
   90&     83.2&   93&     41&     86.6&   84.6&   88.2&   89.8 \\
$\{ {}^{6}\vect{z}(n) \}$ &     0&  37.4&   0.2&    0&  0.2&    58.2&   77.6&   
80.8&   94.6&   56.2&   78.8&   7.8&    27.8&   14.4&   25.4 \\
$\{ {}^{7}\vect{z}(n) \}$ &     0&  2.8&    2.2&    6.4&    0&  96.4&   93.6&   
76.4&   91.4&   30&     30.2&   37.8&   55.2&   36.4&   82.2 \\
$\{ {}^{8}\vect{z}(n) \}$ &     0&  95.2&   5.2&    0.8&    1&  97.2&   98.2&   
92.4&   85.6&   86.6&   94.4&   83.2&   65.2&   59.8&   89.2 \\
$\{ {}^{9}\vect{z}(n) \}$ &     0.2&    87.8&   80.2&   76&     0.2&    96.8&   
99.4&   87.4&   89&     87.8&   12.4&   89.4&   83.4&   77.8&   89.2 \\
$\{ {}^{10}\vect{z}(n) \}$ &    0&  92.4&   28.4&   1.2&    2.8&    98.2&   
98.4&   94.8&   85.4&   50.4&   81.8&   84.4&   68.4&   6.2&    4.2 \\
$\{ {}^{11}\vect{z}(n) \}$ &    0&  0&  0.4&    0.6&    5.2&    4&  63.8&   
89.2&   1.6&    60.8&   61.8&   58.8&   49.4&   81&     87.2 \\
$\{ {}^{12}\vect{z}(n) \}$ &    0&  24.8&   1.2&    1.6&    0&  88.6&   95.4&   
89&     1.6&    3.6&    12.2&   20.8&   2.6&    0.6&    86.4 \\
$\{ {}^{13}\vect{z}(n) \}$ &    0&  4.8&    27.2&   0.2&    3.6&    82.8&   
95.6&   81&     7.2&    92.6&   86&     67.2&   67.8&   66.4&   86 \\
$\{ {}^{14}\vect{z}(n) \}$ &    0&  30.6&   7.8&    1.4&    0&  90.6&   95.8&   
78&     83.8&   6.6&    79&     56.2&   68&     42.6&   88.6 \\
$\{ {}^{15}\vect{z}(n) \}$ &    0&  47.2&   1&  0.2&    0&  85&     96.4&   
89.4&   44.6&   85.8&   87.4&   66.8&   36.4&   58.2&   82.2 \\
$\{ {}^{16}\vect{z}(n) \}$ &    0&  14.4&   14.2&   0&  0&  95.4&   96&     94& 
 
   59.4&   76.6&   96.6&   27.6&   9&  15.6&   7.2 \\
$\{ {}^{17}\vect{z}(n) \}$ &    0&  78.4&   1.4&    0.4&    0.2&    83.6&   
92.4&   95&     92.4&   5.4&    28.4&   16.6&   3.4&    1.2&    87.8 \\
$\{ {}^{18}\vect{z}(n) \}$ &    0&  37.6&   2.2&    1.6&    0&  87.6&   96.6&   
86&     86.2&   92.4&   32.8&   16.4&   22.8&   71.8&   86.2 \\

    \botrule
    \end{tabular}
    \caption{Similar to Tab. \ref{subtab:variables_observed}, except that the 
            values in the tables show the success rate 
            Eq.~\eqref{eq:Eobs_average} which only depends on the estimated
            model state variables and the data.
            }
    \label{tab:variables_observed_estim_error}
\end{table*}   
\begin{table*}\footnotesize
    \centering   
    \begin{tabular}{c*{15}{r}}
         \toprule    
         \textbf{True }& \multicolumn{15}{c}{\textbf{Observed state variables $c$}}\\ 
         \textbf{trajectory} & \text{1} & \text{1-2} &  
         \text{1-3} & \text{1-4} & \text{1-5}
         & \text{1-2-3} & \text{1-2-4} & \text{1-2-5} & \text{1-2-6} 
         & \text{1-2-7} & \text{1-2-8} & \text{1-3-5} & \text{1-3-6}
         & \text{1-3-7} & \text{1-4-7} \\       
        \colrule
$\{ {}^{1}\vect{z}(n) \}$ &     0&  0&  0.2&    50.4&   0&  0.2&    0&  90&     
1.2&    84.2&   0&  52.8&   88.8&   91.6&   90.4 \\
$\{ {}^{2}\vect{z}(n) \}$ &     0&  10.4&   1.2&    23.4&   3&  3.6&    91.6&   
0.4&    19&     74.8&   66.8&   86&     59&     19.6&   71.2 \\
$\{ {}^{3}\vect{z}(n) \}$ &     0.2&    37.6&   22.4&   4.6&    28&     66.6&   
95&     89.8&   71.6&   98.8&   31.2&   21&     87.2&   95.4&   95.6 \\
$\{ {}^{4}\vect{z}(n) \}$ &     0&  1.8&    0.6&    8.4&    11.6&   78.2&   
95.2&   93.8&   27.8&   96.4&   85&     66&     79.4&   31.2&   94.8 \\
$\{ {}^{5}\vect{z}(n) \}$ &     0&  67.2&   63.4&   0.8&    5.4&    92.2&   
96.6&   90&     84.2&   93.4&   0.6&    92.4&   90&     92&     92.8 \\
$\{ {}^{6}\vect{z}(n) \}$ &     0&  37.4&   0.6&    1.2&    4.8&    60.2&   78&  
   82.2&   98.6&   57.8&   86&     16.8&   29.8&   38.6&   26.4 \\
$\{ {}^{7}\vect{z}(n) \}$ &     0&  0&  2.2&    7.4&    0.6&    1&  94.2&   4.2& 
   1&  37.6&   30.4&   41.2&   57&     36.6&   82.6 \\
$\{ {}^{8}\vect{z}(n) \}$ &     0&  0&  0&  2.2&    3&  0.6&    0.6&    1.4&    
0&  86.8&   0&  86.4&   0&  61.8&   96.2 \\
$\{ {}^{9}\vect{z}(n) \}$ &     0.2&    94&     83.4&   82&     1.6&    0&  
99.4&   87.4&   91.4&   98&     88.4&   91.4&   88.8&   93.8&   96.4 \\
$\{ {}^{10}\vect{z}(n) \}$ &    0&  0&  58.8&   1.2&    0&  98.2&   98.4&   0&  
1.2&    0&  0&  0&  0.8&    52.4&   92.4 \\
$\{ {}^{11}\vect{z}(n) \}$ &    0&  3.6&    2&  2.2&    18.8&   5.4&    67.2&   
91.2&   23.2&   61.6&   64&     85.6&   51.8&   86.2&   91.6 \\
$\{ {}^{12}\vect{z}(n) \}$ &    0&  55.4&   36&     7.2&    0.4&    90.4&   
96.4&   92.4&   20.8&   5.4&    25.8&   61&     80.6&   15.8&   94.4 \\
$\{ {}^{13}\vect{z}(n) \}$ &    0&  0.2&    38.2&   3&  28.6&   83&     1&  1.2& 
   0.8&    0&  0&  68.6&   77.8&   78.4&   2.8 \\
$\{ {}^{14}\vect{z}(n) \}$ &    0&  31.4&   1.4&    4.4&    0.2&    0&  97.4&   
1.8&    83.8&   7.4&    84.4&   5.6&    68.2&   94.8&   93.4 \\
$\{ {}^{15}\vect{z}(n) \}$ &    0.2&    70.6&   29.2&   11.4&   4.4&    99.8&   
99.2&   98.8&   97.8&   99.8&   96.4&   72.6&   57.6&   92.2&   94 \\
$\{ {}^{16}\vect{z}(n) \}$ &    0&  0.2&    0&  0&  0&  95.4&   0&  0&  0.2&    
0&  0&  2&  15.6&   0&  0.6 \\
$\{ {}^{17}\vect{z}(n) \}$ &    0&  78.4&   2&  10&     0.4&    0&  0&  0.2&    
1.4&    0&  0&  50.2&   3.4&    43.8&   2.8 \\
$\{ {}^{18}\vect{z}(n) \}$ &    0&  43.2&   17.4&   46.4&   11.2&   91&     
97.2&   97.8&   94&     98.4&   92.2&   58.6&   83.6&   92.8&   92 \\
    \botrule
    \end{tabular}
    \caption{This table show the statistic of the prediction error for initial 
            guess method \ref{enum:iguess1}. The 
            table the table has to be interpreted in the same way as Tab.      
            \ref{subtab:variables_observed}. In contrast to Tab. 
            \ref{subtab:variables_observed} the numbers show $\left< 
            {}^i\mathrm{PE}^c \right>$, Eq.~\eqref{eq:PE_average},
            which is the percentage
            of successful predictions by considering the prediction error
            ${}_h^i\mathrm{PE}^c$, Eq.~\eqref{eq:PE}. An estimation 
            is considered as successful if ${}_h^i\mathrm{PE}^c < 0.5$.
            The length of the prediction window is $N_\mathrm{pred} = 300$   
            and time steps of length $\Delta t = $ are used. }
    \label{tab:PE}
\end{table*}   
Using the initial guess method \ref{enum:iguess2}  (exact solutions of the model) 
in \ref{sec:l96_state_par_estimation_initial_guesses} one can create a similar 
statistic (not shown here). We found that the success rate for the state
variables and parameter estimation is almost zero in many if not most cases 
for all combinations of observed state variables and all true trajectories, i.e.
method \ref{enum:iguess2} gives worse success rates compared to 
method \ref{enum:iguess1} (uniformly distributed samples in a box).

As discussed in Sec. \ref{sec:l96_state_par_estimation_initial_guesses} 
using initial guess method \ref{enum:iguess3} (samples close to the true solution) 
is usually not applicable in a real world estimation process, because the true 
trajectories are usually not given. We use it here to estimate the basin size around the 
true trajectories and it turns out that initial guesses uniformly sampled in a 
``tube'' around the true trajectories with a radius of 15 (which is larger than 
the amplitude of the oscillations) 
provide correct estimates with a very high success rate. 
This means that the optimal solution is not only locally observable 
but possesses a basin of considerable size. However, this basin is bent/curved 
in a 
very high dimensional space.

\subsubsection{Prediction Error}
In contrast to considering $\left< {}^iE^c \right>$ and $\left< {}^i\hat{p}^c 
\right>$ only, we also consider the success rate of the prediction, $\left< 
{}^i\mathrm{PE}^c \right>$, Eq.~\eqref{eq:PE_average}. For initial guess method 
\ref{enum:iguess1} the prediction success rate $\left< {}^i\mathrm{PE}^c \right>$ is shown in 
Tab.~\ref{tab:PE}. Remember, that ${}_h^i\mathrm{PE}^c$ can be computed using 
the solution from the estimation process and the measured data $\{ 
{}^i\vect{\eta}^c(n) \}$ only, provided data for $N \leq n \leq N + 
N_\mathrm{pred}$ are available.
The prediction was computed for $N_\mathrm{pred} = 300$ time steps. Here, due to 
noise in the data, an estimate of the model state variables is considered as 
successful if ${}_h^i\mathrm{PE}^c < 0.5$. Note that ``successful''' here does 
not necessarily mean that the prediction of unobserved state variables is accurate 
nor that in the estimation window $n \in [0, \dots, N]$ the observed and 
unobserved model variables and the model parameter are estimated correctly (in 
the sense that ${}_h^iE^c$ is small and ${}_h^i\hat{p}^c \in [8.16, 8.18]$).
One can see that even for two measured variables there are many
combinations of $\{ ^i\vect{z}(n) \}$ and the measured variables
with a large $\left< {}^i\mathrm{PE}^c \right>$ showing successful predictions 
of observed variables. 
Furthermore, when only a single variable is measured the predictions fail for almost all 
true trajectories as shown by $\left< {}^i\mathrm{PE}^\mathrm{(1)} \right> 
\approx 0\%$. These results are consistent with the results obtained from 
Tab. \ref{subtab:variables_observed}, although on average the percentages have 
smaller numerical values. Nevertheless, there are cases where $\left< 
{}^i\mathrm{PE}^c \right>$ is large and $\left< {}^iE^c \right>$ is small 
(example: $c=\mathrm{(1-2)}$, $i=3$) and vice versa (example: 
$c=\mathrm{(1-3)}$, $i=1$). For both cases estimation and prediction examples 
are shown in Fig.~\ref{fig:prediction_error} left 
column (${}_{385}^3E^\mathrm{(1-2)} > {}_{385}^3\mathrm{PE}^\mathrm{(1-2)}$)
and Fig. \ref{fig:prediction_error} right
column (${}_{140}^1E^\mathrm{(1-3)} < {}_{140}^1\mathrm{PE}^\mathrm{(1-3)}$). 
This means that the correlation between $\left< {}^iE^c \right>$ and 
$\left< {}^i\mathrm{PE}^c \right>$ is strong but not perfect. A good prediction 
does not necessarily mean a good estimation during the estimation window. 
It rather \textit{indicates} that if the prediction error is small then
the estimation of unobserved state variables and parameters is good.

Using the initial guess method \ref{enum:iguess2}  (exact solutions of the 
model), we found 
that the success rate $\left< {}^i\mathrm{PE}^c \right>$ is almost zero in 
many if not most cases for all combinations of observed variables and all true 
trajectories, i.e. method \ref{enum:iguess2} gives worse success rates compared 
to method \ref{enum:iguess1} (uniformly distributed samples in a box). The same 
was observed when considering $\left< {}^iE^c \right>$.
A possible explanation for this observation is the fact that, for exact 
solutions,
the term $C_1$ (Eq. \eqref{eq:ts_model_diff})  in  the cost function $C$ is the only
term significantly different from zero, such that the initial values result in a relatively 
small value of the total cost function 
and this may increase the probability to be close to (and kept in) a local 
minimum.

When initial guess method \ref{enum:iguess3} (samples close to the true 
solution) was used, we observed that most success rates of the prediction are 
close to $\left< {}^i\mathrm{PE}^c \right> \approx 100\%$. Nevertheless, there 
are 
also combinations of a true trajectory and measured state variables with a success 
rate close to zero (especially for one and two measured variables) although 
corresponding success rates  $\left< {}^iE^c \right>$ are high. The 
most likely reason is the chaotic dynamics of the model and 
therefore the fast divergence from the data when computing the predictions.

\section{Discussion and Conclusion}  \label{sec:discussion_and_conclusion}
%
Using a chaotic 9-dimensional Lorenz-96 model as a prototypical example we 
studied observability of all its 9 state variables $x_i$ and the fixed model parameter $p$
using different multivariate time series consisting of one to three observables.
Local observability was characterized by a recently introduced measure of uncertainty $\nu_i$
given in Eq.~\eqref{eq:nu}. This analysis indicates that all state variables and the parameter 
can be reconstructed, even in cases where only a univariate time system is available.
It turned out that on average the values of $\nu_i$ for  unmeasured state variables 
are minimal for a delay time $\tau$  between $\tau=0.11$ and $\tau=0.21$ 
(see Fig.~\ref{fig:fig1_lorenz96_nu_vs_tau}). This is in agreement with results reported 
in Ref.~\cite{REKASBP14} where $\tau \approx 0.1$ was found to be an 
appropriate delay time to synchronize a Lorenz-96  model to an observed time 
series using a delay coordinates based coupling scheme. 
Histograms of the uncertainties $\nu_i$ of the fixed model parameter and 
the measured and unmeasured state variables look similar, independent of the number 
of measured variables.
This means that the successful reconstruction of the state $\vect{x}(t)$ and the parameter $p$  
should not depend on the number of measured  state variables in a (multivariate) time series, 
provided 
one initializes the  estimation algorithm close enough to the true solution
(note that the observability analysis presented in 
Sec.~\ref{sec:local_observability} is only locally valid).

In Ref.\cite{REKASBP14} we showed that for 
the Lorenz-96 model synchronization to the data is indeed possible with 
only a single 
measured state variable, only, using a synchronization scheme based on delay vectors of 
the data time series. Hence this result is  in coincidence 
with the fact that the uncertainty values $\nu_i$ are relatively small already for univariate time series
from the Lorenz-96 system.

Furthermore, we addressed the question whether the estimation of the model states is 
also possible if an estimation algorithm is initialized further away from the true trajectory
of the dynamical system underlying the data. To probe this global convergence 
a 
statistical test was performed where an optimization based state and parameter estimation 
algorithm \cite{SBP11} was initialized with different initial guesses for the entire trajectory and 
the model parameter. Three  different methods for generating the initial guesses were used 
(see Sec. \ref{sec:l96_state_par_estimation_initial_guesses}) and compared. 

With method \ref{enum:iguess1} initial guesses were chosen uniformly distributed  in 
a box. With this preparation of initial guesses of the optimization algorithm state and 
parameter estimation in the 9-dimensional Lorenz-96 model was possible with a very 
high success rate if multivariate times series with (at least) three observables are available, while
for two measured state variables, only a fraction of estimation runs was successful 
(see success rates summarized in Tabs.~\ref{subtab:variables_observed} and \ref{subtab:parameter_observed}).  
Note, that the initial guesses generated by method \ref{enum:iguess1} are 
typically far off the trajectory underlying the data. As a consequence state and parameter estimation 
based on univariate time series failed in most cases. Therefore, for practical application
local observability is a necessary but not a sufficient feature of the given 
estimation problem. 

Furthermore, it was shown that an error definition based on the difference 
between the estimated solution of the model variables and the noise free true 
trajectory (of all variables), Eq.~\eqref{eq:A}, gives comparable success rates 
as an error definition based on the difference between the measurement function 
and the data, Eq.~\eqref{eq:Eobs}. For the latter, a lower boundary 
was derived which is valid for a smooth solution. Note, that in all simulations 
the model equations have no errors. The question of whether both error 
definitions would give comparable results if errors in the model equations are 
present, was not addressed.

Using exact trajectories (not coinciding with the true trajectory underlying the data) as initial 
conditions (method \ref{enum:iguess2}) turned out to result in very poor estimation results.
 Hence, initializing the estimation algorithm with an arbitrary solution of the model equations is a 
disadvantage compared to random initial guesses.

High success rates (close to 100\%) were obtained using initial guess method \ref{enum:iguess3} where the 
estimation algorithm is initialized with samples close to true solutions. These results are consistent  
with the low uncertainty observed in the local observability analysis. 
Note, however, that usually this initialization method can not be 
applied with real world data, because the true trajectories used to generated the initial guess 
are typically unknown.

In addition to  considering the success rate of the estimation, 
Eq.~\eqref{eq:A_average}, which can only be computed if the clean trajectories 
of \textit{all} state variables are known (often only one variable can be measured), 
the success rate of prediction, Eq.~\eqref{eq:PE_average}, was considered. 
This prediction error is more suitable for real world applications, 
because it can be computed based on measured data and the 
estimated model state variables and does not require further information about the 
dynamics. In the example considered here a correlation between the 
prediction error and the success rate of the estimation was observed indicating 
that the prediction error is a good measure for the success of the estimation procedure. 
Nevertheless, it was also shown that a small estimation error does not necessarily mean a small 
prediction error and vice versa.

Our results indicate that successful state and parameter estimation crucially 
depends on the selection of available observables (univariate vs. multivariate), 
on the trajectory segment underlying the time series (i.e., the region of the 
state space the trajectory visits during measurements), and 
last but not least, the initialization of the estimation algorithm. The first 
two aspects are typically determined by and during the measurement (or 
experiment) and cannot me changed afterwards. Only the choice of the estimation 
method and of its initialization is (typically) in the hand of the person who is 
analyzing the data. Using a representative algorithm from the 
class of optimization based methods (similar to 4D-VAR) we 
demonstrated that the success may crucially depend on a proper  choice of 
initial guesses. Finding suitable criteria and initialization strategies is thus 
an important open task for future research on state and parameter estimation 
algorithms.

\section*{Acknowledgements}
The research leading to the results has received 
funding from the European Community's Seventh Framework
Programme FP7/2007-2013 under grant agreement 17
No. HEALTH-F2-2009-241526, EUTrigTreat.
S.L. and U.P. acknowledge support from the  
BMBF (FKZ031A147, GO-Bio),
the DFG (SFB 1002) , and the German Center for Cardiovascular Research (DZHK e.V.).

\section*{Appendix: Jacobian matrix of the delay coordinates map} 
\label{app:DG}
%
The Jacobian of the delay reconstruction map Eq.~\eqref{eq:G} with respect to 
$\vect{x}(t)$ and $\vect{p}$ is given by 
\begin{widetext}
\begin{equation} 
  \D_{\vect{x},\vect{p}} \vect{S} (\vect{x}(t), \vect{p}) = 
    \left(    
    \begin{matrix}  
        \D _{\vect{x}} \vect{h} ( \vect{x}(t)) &  \vect{0}    \\
        \D_{\vect{x}} \vect{h}( \phi^\tau (\vect{x}(t), \vect{p}))    
                    \cdot   \D_{\vect{x}} \phi^\tau ( \vect{x}(t), \vect{p})  
        & \D_{\vect{x}} \vect{h}( \phi^\tau (\vect{x}(t), \vect{p}) )) 
                    \cdot   \D_{\vect{p}}\phi^\tau ( \vect{x}(t), 
                        \vect{p}) \\
        \D_{\vect{x}} \vect{h}( \phi^{2\tau} (\vect{x}(t),\vect{p}))    
                    \cdot   \D_{\vect{x}} \phi^{2\tau} ( \vect{x}(t), \vect{p}) 
 
        & \D_{\vect{x}} \vect{h}( \phi^{2\tau} (\vect{x}(t),\vect{p}))     
                        \cdot \D_{\vect{p}}\phi^{2\tau} 
                        ( \vect{x}(t), \vect{p}) \\      %
        \vdots  &  \vdots  \\
        \D_{\vect{x}} \vect{h}( \phi^{(K-1)\tau} (\vect{x}(t),\vect{p}))  
                    \cdot   \D_{\vect{x}}\phi^{(K-1)\tau} (\vect{x}(t),\vect{p}) 
 
        & \D_{\vect{x}} \vect{h}( \phi^{(K-1)\tau} (\vect{x}(t),\vect{p})) 
            \cdot   \D_{\vect{p}}\phi^{(K-1)\tau} (\vect{x}(t),\vect{p})     
    \end{matrix}   \right)                                    
    \label{eq:DG_xp}
\end{equation}
\end{widetext}
where
\begin{align} 
    \D_{\vect{x}} \vect{h}(\vect{\phi}^{\tau'} (\vect{x}(t), \vect{p})) = &
    \left.
    \left(
    \begin{matrix}
        \frac{\partial h_1}{ \partial x_1} & \ldots & \frac{ \partial 
           h_1} {\partial x_M} \\
            \vdots & & \vdots \\
            \frac{\partial h_R}{ \partial x_1} & \ldots & \frac{ 
            \partial h_R} {\partial x_M}
    \end{matrix}
    \right) \right|_{\vect{x}(t), \vect{p}} 
    \notag \\
%
    %
    \D_{\vect{x}} \vect{\phi}^{\tau'} (\vect{x}(t), \vect{p}) = &
    \left.
    \left(
    \begin{matrix}
        \frac{\partial \phi_1^{\tau'}}{ \partial x_1} & \ldots & \frac{ 
            \partial \phi_1^{\tau'}} {\partial x_M} \\
            \vdots & & \vdots \\
            \frac{\partial \phi_D^{\tau'}}{ \partial x_1} & \ldots & \frac{ 
            \partial \phi_D^{\tau'}} {\partial x_M}
    \end{matrix}
    \right) \right|_{\vect{x}(t), \vect{p}}    
    \notag \\
    %
    \D_{\vect{p}} \vect{\phi}^{\tau'} (\vect{x}(t), \vect{p}) = &
    \left.
    \left(
    \begin{matrix}
        \frac{\partial \phi_1^{\tau'}}{ \partial p_1} & \ldots & \frac{ 
            \partial \phi_1^{\tau'}} {\partial p_P} \\
            \vdots & & \vdots \\
            \frac{\partial \phi_D^{\tau'}}{ \partial p_1} & \ldots & \frac{ 
            \partial \phi_D^{\tau'}} {\partial p_P}
    \end{matrix}
    \right) \right|_{\vect{x}(t), \vect{p}} \, ,
    \label{eq:jac_h_phi}    
\end{align}
with $\tau' = 0, \tau, 2\tau, \dots, (K-1)\tau$.
To compute the Jacobian matrix $\D \vect{S}_{\vect{x}, \vect{p}} (\vect{x}(t), 
\vect{p})$  \eqref{eq:DG_xp}
of the delay coordinates map $\vect{S}(\vect{x}(t), \vect{p})$ we have to 
compute 
the Jacobians  \eqref{eq:jac_h_phi}
where $\D_x\phi^{\tau'}  (\vect{x}(t),\vect{p})$  and 
$\D_p\phi^{\tau'} (\vect{x}(t),\vect{p})$ contain derivatives of the flow 
$\phi^{\tau'}$ generated by the dynamical system \eqref{eq:ode_model} with 
respect to state variables $x_j$ and parameters $p_j$, respectively.
The $D \times D$-matrix  $\D_{\vect{x}} \phi^{\tau'} (\vect{x}(t), \vect{p})$  
can be computed by solving the linearized dynamical equations in terms of a 
matrix ODE
\begin{align} 
    \label{eq:lineq}
    \frac{\diff{}}{\diff{\tau}} \mat{Y}(\tau) &= 
            \D_{\vect{x}}\vect{F} \left( \phi^{\tau}(\vect{x}(t), \vect{p}),
            \vect{p} \right) \cdot \mat{Y}(\tau) 
\end{align}
%
where $\phi^{\tau}(\vect{x}(t),\vect{p}) $ is a solution of 
Eq.~\eqref{eq:ode_model} with initial value $\vect{x}(t)$ and
$\mat{Y}(\tau)$ is an $D \times D$ matrix that is initialized as 
$\mat{Y}(\tau=0) = \mathbb{1}_D$, where $\mathbb{1}_D$ denotes the $D \times D$ 
identity matrix.
Similarly, the $D \times P$-matrix  $\D_{\vect{p}}\phi^{\tau} (\vect{x}(t), 
\vect{p})$ is obtained as a solution of the matrix ODE  \cite{Kawakami}
\begin{align}    
    \frac{\diff{}}{\diff{\tau}} \mat{Z}(\tau) = &
        \D_{\vect{x}}\vect{F} \left( \phi^{\tau}(\vect{x}(t), \vect{p}),
            \vect{p} \right) \cdot \vect{Z}(\tau) \notag \\
        & + \D_{\vect{p}}\vect{F}\left( \phi^{\tau}(\vect{x}(t), \vect{p}),
        \vect{p} \right)
    \label{eq:lineq2}        
\end{align}
with $\mat{Z}(\tau=0)=0$. $\D_{\vect{x}}\vect{F} (\dots)$ and 
$\D_{\vect{p}}\vect{F} (\dots)$ denote the Jacobians containing 
derivatives $\partial F_i(\dots) / \partial x_j$ and $\partial F_i(\dots) / 
\partial p_j$, respectively.
Solving \eqref{eq:lineq} and \eqref{eq:lineq2} simultaneously with the system  
ODEs 
\eqref{eq:ode_model}  we can compute 
\begin{align}
    \D_{\vect{x}}\phi^\tau (\vect{x}(t), \vect{p}) = & \vect{Y}(\tau) \\
    \D_{\vect{x}}\phi^{2\tau} (\vect{x}(t), \vect{p}) = &\vect{Y}(2\tau)\\
    \vdots \notag \\
    \D_{\vect{p}}\phi^\tau (\vect{x}(t), \vect{p}) = & \vect{Z}(\tau) \\
    \D_{\vect{p}}\phi^{2\tau} (\vect{x}(t), \vect{p}) = & \vect{Z}(2\tau) \\
    \vdots \notag
\end{align}
and use these matrices to obtain the Jacobian matrix $\D_{\vect{x},\vect{p}} 
\vect{S} (\vect{x}(t), \vect{p})$ Eq.~\eqref{eq:DG_xp} of the delay coordinates
map $\vect{S}$ Eq.~\eqref{eq:G}.                

\bibliographystyle{elsarticle-num}
\bibliography{SBPAKREL_arxiv}

\end{document}